\documentclass[journal]{IEEEtran}
\usepackage{stmaryrd}
\usepackage{amsmath}
\usepackage{amssymb}
\usepackage{psfrag}
\usepackage{epstopdf}
\usepackage{todonotes}
\usepackage{graphicx}
\usepackage{cite}
\usepackage{algorithmic}
\usepackage{algorithm}
\usepackage{svg}
\usepackage{mathtools}
\usepackage{hyperref}
\usepackage{booktabs}
\usepackage{hyphenat}
\usepackage{multirow}
\usepackage{booktabs}
\usepackage{textcomp}
\usepackage{accents}
\usepackage{minibox}
\usepackage{fancyhdr}
\usepackage{kantlipsum}
\fancypagestyle{fancy}

\fancypagestyle{plain}{
  \fancyhf{}
  \fancyhead[C]{Conference on \LaTeX}     %% C or L or R.
  \fancyfoot[L]{This is a notice}%                        %% C or L or R.

}
\usepackage{eso-pic}

\usepackage[caption=false]{subfig}

\DeclareMathOperator*{\argmin}{arg\,min}

\usepackage{float}
\begin{document}
% \title{{Machine Learning-assisted Bayesian Inference for Near Real-Time Jamming Detection and Classification in 5G New Radio (NR)}}

\AddToShipoutPictureBG*{%
  \AtPageUpperLeft{%
    \hspace*{\dimexpr0.175\paperwidth\relax}%%  change \dimexpr0.5\paperwidth\relax appropriately
    % \chead{\large This work has been submitted to the IEEE for possible publication. Copyright may be transferred without notice, after which this version may no longer be accessible.}
    % \makebox(0,-0.75)[c]{\large This work has been submitted to the IEEE for possible publication. Copyright may be transferred without notice, after which this version may no longer be accessible.}%
    \minibox[c]{\\ \\ \\ \\ \emph{This work has been submitted to the IEEE for possible publication.} \\ \emph{Copyright may be transferred without notice, after which this version may no longer be accessible.}}
}}
\AddToShipoutPictureBG*{%
  \AtPageLowerLeft{%
    \setlength\unitlength{1in}%
    \hspace*{\dimexpr0.5\paperwidth\relax}%%  change \dimexpr0.5\paperwidth\relax appropriately
    \makebox(0,0.75)[c]%
}}

\title{{Bayesian Inference-assisted Machine Learning for Near Real-Time Jamming Detection and Classification in 5G New Radio (NR)}}
\author{Shashank Jere, Ying Wang, Ishan Aryendu, Shehadi Dayekh and Lingjia Liu
\thanks{S. Jere and L. Liu are with \emph{Wireless@Virgnia Tech}, Bradley Department of ECE at Virginia Tech. Y. Wang and I. Aryendu are with Stevens Institute of Technology. S. Dayekh is with Deloitte \& Touche LLP. The corresponding author is L. Liu (ljliu@ieee.org). Part of this work was presented at the \emph{IEEE} 23rd International Conference on High Performance Switching and Routing (HPSR)~\cite{WangHPSR2022}. This work was supported in part by Deloitte \& Touche LLP's Cyber 5G Strategic Growth Offering. }
}
\maketitle
% \thanks{Part of the work was presented in~\cite{WangHPSR2022} at the 2022 \emph{IEEE} 23rd International Conference on High Performance Switching and Routing (HPSR).}

\begin{abstract}

The increased flexibility and density of spectrum access in 5G New Radio (NR) has made jamming detection and classification a critical research area. To detect coexisting jamming and subtle interference, we introduce a Bayesian Inference-assisted machine learning (ML) methodology. Our methodology uses cross-layer critical signaling Key Performance Indicator data collected on a Non-Standalone (NSA) 5G NR testbed to leverage supervised learning models, and are further assessed, calibrated, and revealed using Bayesian Network Model (BNM)-based inference. The models can operate on both instantaneous and sequential time-series data samples, achieving an Area under Curve above 0.954 for instantaneous models and above 0.988 for sequential models including the echo state network (ESN) from the Reservoir Computing (RC) family, for various jamming scenarios. 
Our approach not only serves as a validation method and a resilience enhancement tool for ML-based jamming detection, but also enables root cause identification for any observed performance degradation. 
The introduced BNM-based inference proof-of-concept is successful in addressing 72.2\% of the erroneous predictions of the RC-based sequential detection model caused by insufficient training data samples collected in the observation period, thereby demonstrating its applicability in 5G NR and Beyond-5G (B5G) network infrastructure and user devices. 

\end{abstract}


\section{Introduction}
\label{introduction}
With the advent of next-generation cellular networks, the available spectrum resources are facing increasing congestion, especially in the sub-6 GHz spectrum.
As government agencies and private sector businesses continue to adopt 5G New Radio (NR) networks for their end services, protecting these networks against malicious attacks has become crucial in preventing Quality of Service (QoS) degradation or worse, link failures in mission-critical applications.
The first step in mitigating and protecting critical communications links against the presence of malicious jamming attacks or undesirable interference signals is accurate and low-latency detection of interference, both known and unknown. Such detection provides sufficient time for proactive defense before potential QoS degradation or link failures. While 5G NR networks offer flexibility for performance enhancements and vertical customization, they also present increased complexity in terms of security.
There is a pressing need to understand the vulnerability of 5G NR and 5G-Advanced systems to various types of jamming and to equip the networks with an intelligent solution that can autonomously detect jamming and continuously learn from past experience. With new technology come cyber attacks often aimed at reliability and latency-centric applications and services. Traditional case-by-case cyber security strategies and methodologies become limited and powerless as the flexibility of network configuration increases significantly, necessitating more robust strategies.

% \textcolor{blue}{(Add new paragraph on need for high performance (high detection and false alarm rate), near real-time and low complexity jamming detection and classification, especially in UAV networks. Addresses R1 Q1).}

Furthermore, with the advent of unmanned aerial vehicles (UAVs) and their planned integration with next-generation wireless systems in various roles, there has been significant research effort to address the ground-to-unmanned Aerial Vehicle (UAV) (G2U) communication network with an aim to minimize end-to-end transmission delay in the presence of interference in Space-Air-Ground Integrated Networks (SAGIN)\cite{arya2023ground, rock2023radar, hosseinalipour2020interference}. The algorithm presented in this work for detecting and classifying jamming and interference contributes to the low latency G2U and  UAV-to-UAV (U2U) services being used in practice by a network controller as a system parameters selection criteria, where the preference and set of parameters given the knowledge of the existence and type of jamming and interference leading to the lowest transmission latency, can be incorporated into the transmission when applying Command and Control (C2) standards to mission-critical G2U and U2U services.
% \textcolor{blue}{(Add some motivation for this work related to O-RAN xApps).} 
In addition, the Open Radio Access Network Alliance (O-RAN)~\cite{O-RANAlliance2018O-RAN:RAN} which is an industry-wide effort, has put forth a new paradigm for RAN design namely, Open RAN (O-RAN) that advocates for decentralization of the various components in a RAN, thereby allowing for reconfigurability, interoperability and development of new optimization algorithms via intelligent and data-driven controllers. 
A key foundational principle within O-RAN is that of RAN Intelligent Controllers (RICs) which can execute closed-loop optimization routines to orchestrate the RAN~\cite{Bonati2021, polese2022understanding}.
Specifically, the near-real-time (near-RT) RIC, which operates at the network edge and communicates with multiple RAN nodes has been specified to run control loops with a time scale of $10$ ms to $1$ s~\cite{ORAN_nearRT_RIC_spec}.  
The near-RT RIC executes custom algorithms called xApps, which are microservices performing radio resource management~\cite{polese2022understanding}. Incorporating machine learning-based algorithms for closed-loop control of the RAN via the near-RT RIC may become necessary in the future as standardization activities for 6G pick up pace.
This necessitates the development of machine learning-based models that meet the near-RT latency requirements, ideally permitting model training within this period for tasks such as jamming or interference detection and taking the requisite control action to mitigate performance degradation.

\subsection{Related Work}
\label{sec:Related_Work}
In this section, we provide an overview of the current state-of-the-art in RF anomaly and network intrusion detection. We focus primarily on identifying anomalous signals, including jamming and interference signals of known protocol types, which is the primary focus of our work.
The emergence of new technologies such as LTE License Assisted Access (LTE-LAA)~\cite{LTE-LAA_Lee2016, RevistingLTELAA2021} in 3GPP (3rd Generation Partnership Project) Release 13 has raised concerns about the fairness of unlicensed spectrum usage by other technologies such as WiFi. LTE-LAA uses a CSMA/CA-based channel access methodology to promote fair spectrum usage~\cite{LTE-WiFi_Chen2017,Fairness_Massimiliano2018}. Similar co-existence issues have carried over to 5G NR, particularly in the C-band where fixed satellite broadcasts occur in the upper frequencies. This frequency range is also close to the CBRS band, which has great flexibility in terms of user priority, making it challenging to ensure fair spectrum usage.
Given such co-existence issues, it is crucial to develop an accurate and latency-aware strategy for jamming detection.
Traditional spectrum sensing algorithms, such as energy detection~\cite{EnergyDet_Shankar2005, EnergyDet_Yuan2007, EnergyDet_Ghasemi2007}, waveform-based sensing~\cite{WvfmSensing_Tang2005, WvfmSensing_Sahai2006}, and matched-filtering sensing~\cite{MatchedFiltering_Tandra2005} have limitations that make them difficult to adapt to rapidly changing wireless channels and interference environments. These algorithms are highly scenario-dependent, requiring careful tuning, and have several shortcomings. For instance, they cannot detect jamming or interference signals with lower received strength than the average signal strengths of the legitimate signals. Additionally, they require knowledge of a single interference or jamming source beforehand, which is not practical. Lastly, these methods need the legitimate communication links to be idle in order to detect intrusion signals, which is not possible in congested spectrums such as the C-band or the CBRS band.
To address the limitations of traditional detection methods, data-driven or learning-based approaches to anomaly detection have emerged as a successful alternative. Supervised learning methods require a significant data collection and annotation effort, while unsupervised learning on the other hand offers a useful alternative. Our previous work~\cite{WangHPSR2022} incorporated an auto-encoder ensemble architecture to detect previously unseen jamming attacks while also considering adversarial training, thus overcoming the drawbacks of classical detection methods.

Intrusion detection techniques can be categorized into two types: anticipation and prevention defense (APD) and anomaly-based intrusion detection systems (IDS). APD relies on prior knowledge and probabilities of possible attacks and malfunctions, while IDS focuses on learning expected system responses and detecting anomalies. APD heuristics such as sums of vulnerability scores or missing patches are often criticized for being misleading, whereas evolved APD approaches that use software-defined 5G NR/Beyond-5G (B5G) testbeds, advanced machine learning algorithms, and large computational power provide a deep understanding of the signal under test and the radio environment.
In contrast to APD, several in-band IDS-based anomaly detection techniques have been investigated, such as the unsupervised anomaly detection method proposed in~\cite{Walton2017UnsupervisedTransmissions}, which uses a combination of Long-Short Term Memory (LSTM) and Mixture Density Network (MDN) to analyze temporal data. The dynamic intuitionistic fuzzy sets algorithm introduced in~\cite{IDS_Xie2022} shows superior performance compared to state-of-the-art IDS algorithms. However, the lack of a synthesized understanding of the cross-layer response of the system under test (SUT) to various attacks and upper-layer domain knowledge limits the accuracy and efficiency of anomaly-based IDS.
To address this limitation, Bayesian networks have been applied previously to network diagnosis~\cite{You2009ResearchNetworks} and reliability analysis~\cite{Benrhaiem2020BayesianNetworks}. 
% to enhance anomaly detection and machine learning model transparency by utilizing posterior evidence. 
More recently, jamming detection specifically in 5G NR has garnered research interest, e.g.,~\cite{Ornek2022} introduces a low-complexity jamming detection technique by measuring the error vector magnitude (EVM) in each resource block (RB).
Specifically, detection of jamming specifically on the unencrypted (during initial access) synchronization signal block (SSB) using signal processing-based methods is studied in~\cite{Wang2023}.
In the context of machine learning (ML) approaches, real-time ML-based jamming detection in 5G NR specifically using Hoeffding Decision Tress is investigated in~\cite{Arjoune2020}. 
In~\cite{Li2022}, a software defined radio (SDR) is used to launch jamming attacks against a UAV receiver and conventional machine learning approaches are implemented for jamming detection and classification using extracted radiometric features. It also investigates using convolutional neural networks (CNNs) for classification with spectrogram images as input.
A CNN-based autoencoder approach was introduced in~\cite{Bouzabia2023} for detecting Low Probability of Intercept (LPI) jamming signals specifically in the context of radar communications.

However, all the above works for jamming detection except~\cite{Li2022} are based on simulation and unlike this work, do not leverage cross-layer statistical KPI data collected on a real 5G NR (NSA) testbed. 
Furthermore, the learning-based approaches considered in the above works are typically not amenable to low complexity implementations that permit near real-time training and inference.
On the contrary, in this work, we leverage Reservoir Computing (RC), specifically the echo state network (ESN) for jamming detection and classification.
We empirically show via measured CPU run times that the ESN-based detector and classifier can also be implemented in a near real-time fashion on a 5G NR NSA testbed, allowing near real-time training and inference.
% Additionally, as we show in later sections that our introduced method which leverages Reservoir Computing (RC), specifically the echo state network (ESN) can also be implemented in a near real-time fashion on a 5G NR NSA testbed \textcolor{blue}{,allowing near real-time training and inference}.
Furthermore, we utilize Bayesian causal inference built using the available domain knowledge to validate the ML model performance and demonstrate its utility in improving the accuracy and reliability of the jamming detection performance.
% \textcolor{blue}{Such a combination of a near real-time jamming detection and classification approach that can be implemented with acceptable complexity at both the network edge as well as a mobile UE with accompanying proof-of-concept from data collected on a real 5G NR testbed is yet to be found in state-of-the-art.}
Unlike the work presented in this paper, the state-of-the-art lacks an approach for near real-time jamming detection and classification that is implementable with acceptable complexity at the network edge and on mobile UEs, along with supporting proof-of-concept from a real 5G NR testbed data.
We clearly state our primary original contributions in the following section.
% However, a similar application of Bayesian networks for jamming detection incorporating cross-layer statistical data is missing in state-of-the-art.
% In this study, the output of the models is fed back to the Bayesian model to determine the jamming types, effects, and corresponding strategy.

\subsection{Primary Contributions}
\label{sec:contributions}
The main contributions of this work are summarized below:
\begin{itemize}
    % \item A supervised learning approach to interference detection is introduced based on the physical layer (PHY) and higher layer KPI data on a real 5G non-standalone (NSA) testbed.
    \item We introduce a supervised learning approach for jamming detection and classification in a cellular communications system that leverages both physical layer (PHY) and higher layer key performance indicator (KPI) data, collected using a real 5G NR NSA testbed.    
    
    % \item A systematic strategy that is adapted to various interference scenarios differing in frequency bands and power levels is developed by leveraging instantaneous discriminative models and sequential time series-based  models for interference detection. 
    \item We develop a systematic strategy for jamming detection and classification that can be adapted to varying jamming scenarios with different frequency bands and power levels. This is achieved by utilizing a combination of instantaneous discriminative models and sequential time series-based models, including the echo state network (ESN) from the Reservoir Computing (RC) family.

    \item Through RC, we demonstrate the applicability of a jamming detection and classification approach that delivers high detection accuracy, low false alarm rate while meeting near real-time latency constraints with low computational and memory complexity.
    
    % \item Building on a Bayesian Network Model (BNM), the posterior probability of the presence of interference is evaluated.
    \item We introduce a causality inference approach that leverages a Bayesian Network Model (BNM) constructed using the available domain knowledge in order to improve the transparency of the ML-based jamming detection and classification model.
    
    % \item The results of the Bayesian causal inference have been shown to align with the expected KPI values as per 3GPP standards.
    \item We demonstrate a proof-of-concept of using domain knowledge-driven causal inference via the BNM to detect a jamming signal, given the topology and statistical distribution among various parameters collected via legitimate communications. We show that incorrect predictions in ML-based models due to insufficient data samples can be mitigated via the introduced Bayesian analysis.  
\end{itemize}
The remainder of the paper is summarized as follows. In Sec.~\ref{sec:jamming_problem}, we present the problem formulation for the jamming detection task.
In Sec.~\ref{sec:system_description}, we describe the hardware platform used for this work, followed by elaborating on the supervised learning models utilized. The motivation and strategy for Bayesian causality inference is also presented. The complete simulation results are presented in Sec.~\ref{sec:performance_evaluation}. Finally, we provide concluding remarks and directions for future work in Sec.~\ref{sec:conclusion}.

\section{Problem Formulation}
\label{sec:jamming_problem}
\begin{figure}[htbp]
\centering
\includegraphics[width=0.4875\textwidth]{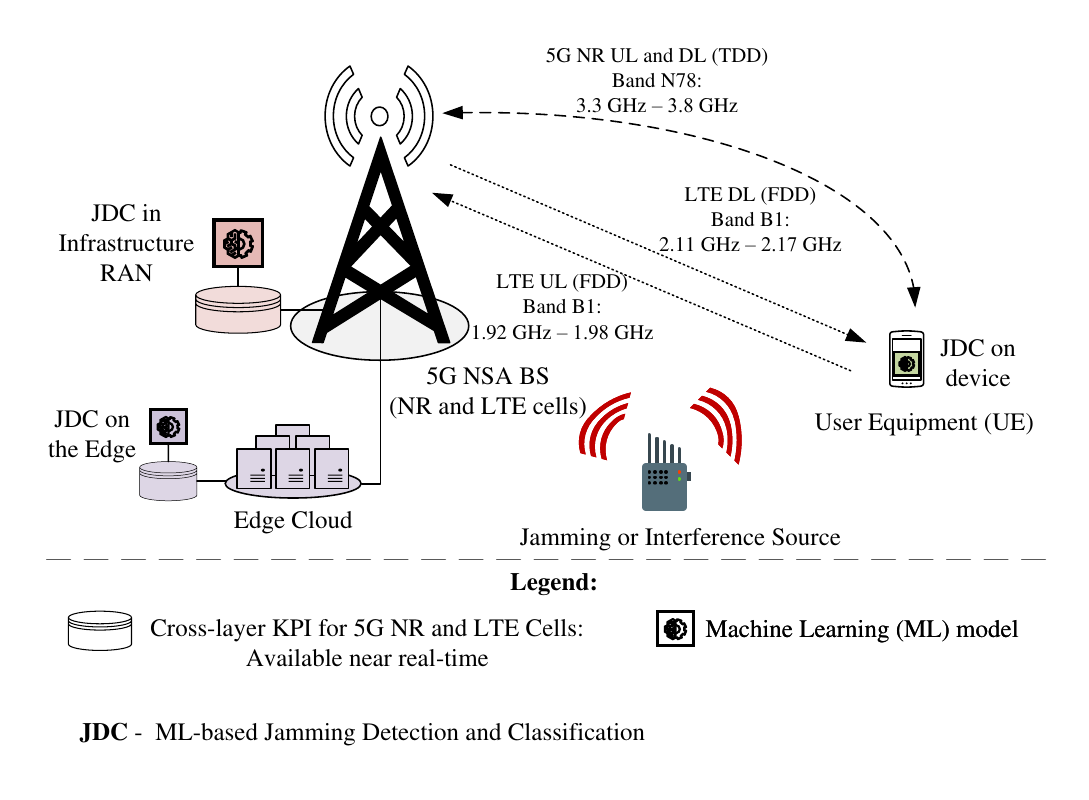}
\caption{System model with an adversary transmitting a jamming signal.}
\label{fig:system_model}
\end{figure}

In the remainder of the paper, we shall use the terms ``jamming'' and ``interference'' interchangeably, noting however, that this is only in the context of our hardware testbed setup and the subsequent machine learning-based detection and Bayesian inference steps.
A jamming detection and monitoring application is designed for a 5G NSA wireless network. 
% as shown in Fig.~\ref{system}.
% A jamming detection and cyber monitoring application of a 5G NSA wireless network as shown in  Fig.~\ref{system} was designed and implemented. 
This involves the steps of: i) jamming signal generation, ii) communication configuration, iii) data logging, iv) intelligent analysis, and v) real-time feedback, all of which are integrated throughout the life cycle of the jamming detection module. 
In the presence of a jamming or a general interference signal, the communication links between the Base Station (BS) and the legitimate users or user equipments (UEs) would be impacted. Statistical information in terms of cross-layer KPIs from the data plane including the physical layer (PHY), Medium Access Control layer (MAC), Radio Link Control layer (RLC), and Packet Data Convergence Protocol (PDCP) are collected and analyzed at the jamming detector module. The cross-layer data and side information available from the network and application layers are also input to the Bayesian Network Model (BNM) for cyber inference analysis. The jamming/interference detector and cyber inference analyzer module can be located at the BS to monitor the statistical information. It can also be a separate node fetching this information from the BS periodically. The former design, i.e., co-located at the BS, complies with the O-RAN~\cite{O-RANAlliance2018O-RAN:RAN} architecture. In the latter setting, the interference detector and analyzer module is located in a separate hardware unit and minimally impacts the BS performance. 
% \textcolor{blue}{Finally, }
Fig.~\ref{fig:system_model} depicts a scenario where a legitimate user, e.g., a UE is affected by the presence of an undesirable interference or a malicious jamming signal.

% Original figure
% \begin{figure}[htbp]
% \centering
% \includegraphics[width=0.75\textwidth]{figures/system_new.png}
% \caption{System Overview}
% \label{system}
% \end{figure}

% \begin{figure*}[h]
% \centering
% \includegraphics[width=0.85\textwidth]{figures/system.png}
% \caption{System Overview}
% \label{system}
% \end{figure*}

% Fig. 1 placeholder

% Fig.2 placeholder
\begin{figure}[htbp]
\centering
\includegraphics[width=0.475\textwidth]{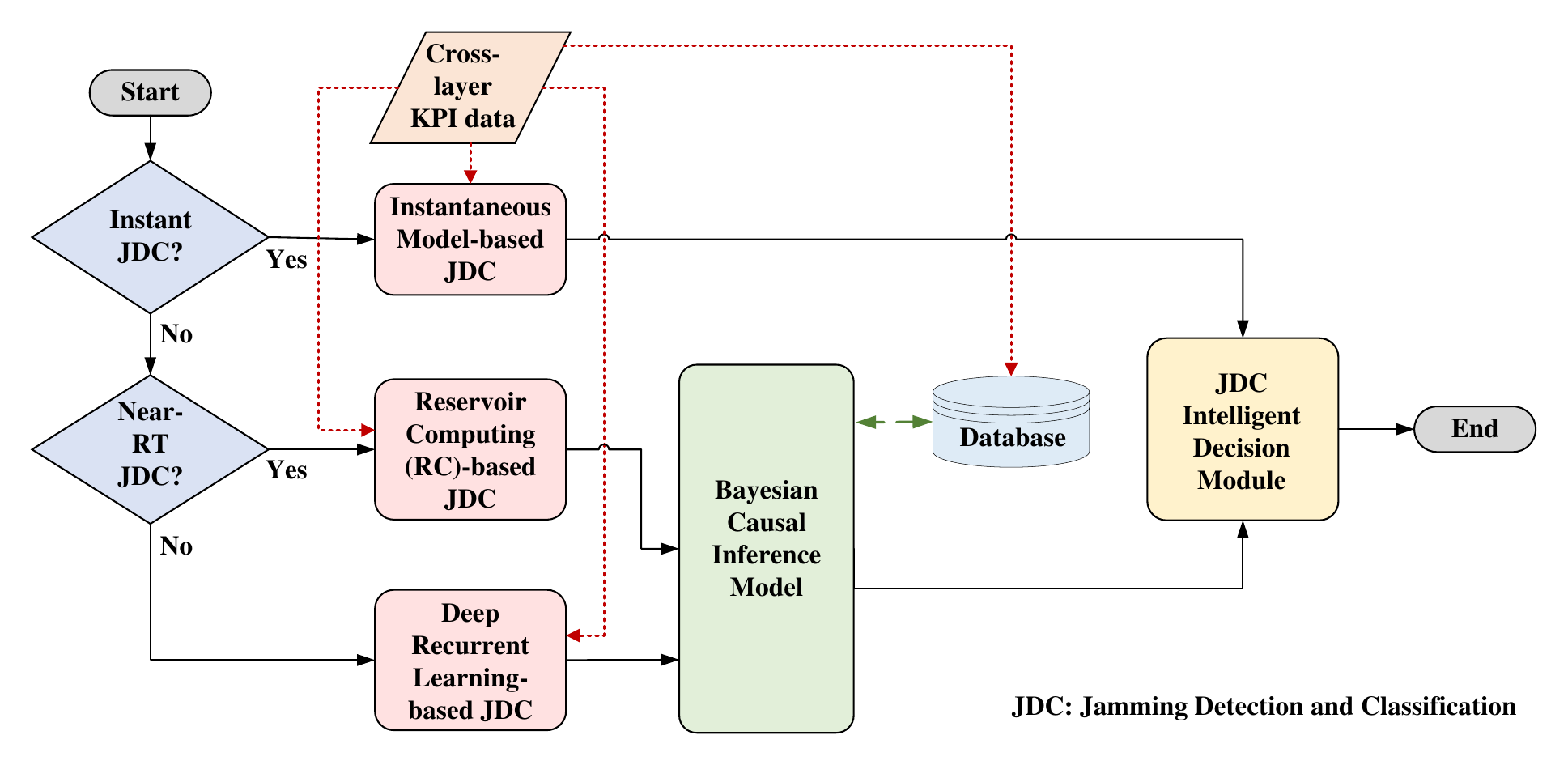}
\caption{Flowchart of the Jamming Detection and Classification Decision Module. }
\label{fig:flowchart}
\end{figure}

In our experimental setup, jamming signals spanning multiple center frequencies and power levels 
% multiple types of jamming signals
are generated and the corresponding statistical KPI data is collected and stored in the implementation platform for the purpose of training various machine learning models and subsequent performance evaluation. The jamming detection problem can be formulated as two hypotheses $H_0$ and $H_1$:
\begin{equation}
H_0: \mathbf{r}[n] = \mathbf{s}[n] + \mathbf{w}[n]; n \in [1, N],\label{eq0}
\end{equation}
\begin{equation}
H_1: \mathbf{r}[n] = \mathbf{s}[n] + \mathbf{w}[n] + \mathbf{j}_m[n]; n \in [1, N], \\
m\in[0, N_{\mathrm{jam}}], \label{eq1}
\end{equation}
where $N$ is the total number of data samples collected and $N_{\mathrm{jam}}$ is the total number of known jamming signal types considered (including changes in power level and center frequency for a jamming signal of a given protocol type).
The discrete time index is denoted by $n$, whereas $m$ is the index denoting the type of jamming signal, with $m=0$ if the type of jamming signal is unknown.
Here, $\mathbf{r}[n] \in \mathbb{R}^{N_f} := [\boldsymbol{r}_{\text{UE}}[n], \boldsymbol{r}_{\text{BS}}[n]]^T$  is  the concatenated vector of the cross-layer KPI data at time index $n$, available at both the UE and the BS in near real-time.
$N_f$ denotes the total number of cross-layer KPI parameters that act as the ``features'' of the data sample at each time step.
$\mathbf{j}_{m}[n] \in \mathbb{R}^{N_f}$  represents the perturbation in the concatenated cross-layer KPI data at time $n$ caused by the presence of a malicious jamming signal or undesirable interference;
$\mathbf{s}[n] \in \mathbb{R}^{N_f}$ is the concatenated cross-layer KPI data which is ``expected'' in the absence of any jamming or interference; $\mathbf{w}[n] \in \mathbb{R}^{N_f}$ represents the combined measurement or environment noise that is added to the cross-layer KPI data regardless of jamming. 
% is the concatenated additive channel environment noise. 
% $n$ is the index of discrete time slots, $m$ is the index of jamming type, where $m=0$ if the jamming type is unknown. 
Furthermore, we can define  
\begin{equation}
\mathbf{Y}[n] = f(\mathbf{r}[n],\mathbf{r}[n-1], ...,\mathbf{r}[n-T]),
\label{eq:temporal_data}
\end{equation}
where $\mathbf{Y}[n] \in \mathbb{R}^{N_f \times T}$ represents a sequence of the cross-layer KPI data collected over $T$ time steps.
% is the collated statistical information from the BS and the UE for the communication link between them being for the current and the past $k$ times slots.
The function $f(\cdot)$ denotes feature-wise normalization that may be performed on the collected raw data before being input to the machine learning (ML) model.
Our approach for jamming detection comprises designing a test statistic $\Lambda(\mathbf{Y})$ and comparing it with a threshold $\eta$, which can be denoted as:
$H_1 = \text{True}$ if $\Lambda(\mathbf{Y}) > \eta$ and  
$H_0 = \text{True}$ if $\Lambda(\mathbf{Y}) \leq \eta$.
The test statistic $\Lambda(\mathbf{Y})$ and the threshold $\eta$ are determined by the machine learning model being considered.
% Two types of models are proposed and compared: supervised discriminative models and unsupervised anomaly detection models.
In the supervised learning-based discriminative model, the objective is to maximize $\Pr(\Lambda(\mathbf{Y}) > \eta|H_1)$ and maximize $\Pr(\Lambda(\mathbf{Y}) < \eta|H_0)$.
The flowchart describing the overall jamming detection module is shown in Fig.~\ref{fig:flowchart}.
% \textcolor{blue}{Add a short description of the algorithm flowchart in Fig. 2.}
Based on the jamming attack scenario and the deployed hardware configuration, the decision-making process is split into two branches as shown in Fig.~\ref{fig:flowchart}: a) whether an instantaneous jamming detection decision is required, or b) jamming detection as well as classification is required and there are no real-time constraints.
For branch (b), near real-time constraints ($10$ ms to $1$ sec) necessitate the use of the introduced Reservoir Computing (RC)-based detection and classification. On the other hand, if there are no such near-RT constraints, the algorithm flow proceeds to deploy a deep recurrent learning model (e.g., LSTM, GRU, CNN-LSTM, etc.).
% within $180$ ms detection required or there are no stringent detection time requirements. 
The inputs of the system are the Key Performance Indicator (KPIs) spanning multiple protocol layers with $180$ ms sampling periods that could be collected at the Radio Access Network (RAN) Infrastructure (BS or ``Cell Site''), the Edge cloud, and the UE, including embedded IoT devices. 
The instantaneous jamming detection model takes as input a single time step of the KPI parameter data to perform a binary detection decision, while the sequential models, including LSTM and ESN require a longer sequence of inputs and a larger number of training samples to detect the type of interference with high accuracy. 
The cross-layer data collection process is compatible with both O-RAN and non-O-RAN architecture. The Bayesian Causal Inference Model (introduced in Sec.~\ref{sec:causal} and evaluated in Sec.~\ref{sec:bayesian_results}) is used to aid sequential learning, leading to improved accuracy and robustness when the number of training samples available to train the sequential model is limited.

% improves the robustness and accuracy of the model.    
% In the unsupervised anomaly detection model, the objective is to maximize $P((\Lambda(y) > \eta, H_0)\cap (\Lambda(y) < \eta, H_1))$.

\section{System Description}
\label{sec:system_description}
\subsection{Hardware Platform}\label{platform}
\begin{figure*}[htbp]
\centering
\includegraphics[width=0.925\textwidth]{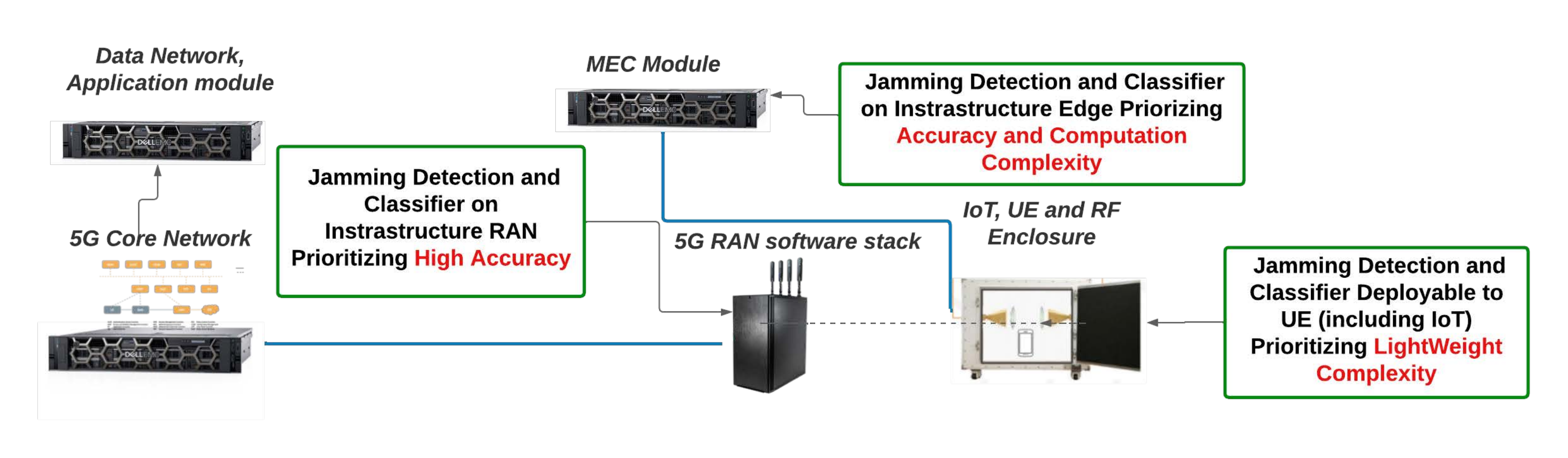}
\caption{System Hardware Platform Setup.}
\label{hardware}
\end{figure*}

The hardware platform used in our implementation of the 5G NSA testbed for experimental data collection is depicted in Fig.~\ref{hardware},
% The implemented hardware platform is depicted in Fig.~\ref{hardware}, 
and consists of the following components: User Equipment (UE), Base Station (BS), Core Network (CN), Cyber Attack Controller (CAC), Jamming Attack Radio Generator (JARG), and a Jamming Detection and Cyber Inference Model (JDCIM). The CAC generates different types of controlled jamming for training and testing purposes. The JARG module emulates the jamming attacks on the legitimate communication link between the UE and the BS inside the RF enclosure. The connections between the BS and the CN, the CAC, and the JARG are via an Ethernet cable. The complete implementation details of this setup can be found in~\cite{wang2021development_1},  and \cite{Wang2021AI-PoweredOptimization}. 
% At the current stage, we assume that the channel noise is additive. 

% Original figure

% \begin{figure*}[h]
% \centering
% \includegraphics[width=0.75\textwidth]{figures/hardware.png}
% \caption{System Hardware Platform Setup}
% \label{hardware}
% \end{figure*}

The jamming waveforms are originally generated in MATLAB and converted to Application Resource Bundle (ARB) files compatible with the Rohde\&Schwarz\texttrademark~ 
SMW2000A Signal Generator. This provides a systematic way to generate physical layer (PHY) signal waveforms for different wireless protocols.  
% Table~\ref{waveform} shows the interference/jamming waveforms that are transmitted at the desired RF center frequency.
In this experiment, we choose the IEEE 802.11n waveform with a bandwidth of $80$ MHz as the baseband signal waveform of choice, which is then transmitted as a jamming signal at the desired center frequency and power level to observe its impact on the legitimate 5G NR NSA cellular communication link between a BS and a UE.
Table~\ref{tab:waveforms_description} summarizes the center frequencies and power levels at which the jamming WiFi waveform is transmitted within the RF enclosure to create the various jamming scenarios.

Considering real-world applications, spectrum utilization, and Federal Communications Commission (FCC) regulations, this paper presents our study conducted in the 1.95 GHz and 2.14 GHz bands for 5G NR NSA traffic, with a specific focus on WiFi jamming or interference detection and classification as a pilot example. This emphasis stems from the recognition that the WiFi signal constitutes the primary source of jamming or interference. The successful classification of various types of piloted WiFi signals underscores the heightened sensitivity of our proposed model. The detection capability of the established model extends beyond the pilot bands and jamming signal types, thereby enabling adaptation to a diverse range of jamming signals.
Note, however, that this choice of IEEE 802.11n WiFi for the waveform is not strict and can be replaced with any other wideband waveform of choice, provided the center frequency and the transmit power can be controlled for the purpose of experimental evaluation.
% \begin{table}[htbp]
% \caption{Jamming Waveforms Description}
% \begin{center}
% \begin{tabular}{|c|c|c|c|c|}
% \hline
% \textbf{Type} & \textbf{Bandwidth}& \textbf{Center Freq}& \textbf{Power} & \textbf{Sample Size}\\
% \hline
% WiFi & 80M & 2.14G & -11dBm & 1135\\
% \hline
% WiFi & 80M & 2.14G & -12dBm &979\\
% \hline
% WiFi & 80M & 2.14G & -13dBm & 884\\
% \hline
% WiFi & 80M & 2.14G & -5dBm & 890\\
% \hline
% WiFi & 80M & 2.14G & 0dBm & 1033\\
% \hline
% WiFi & 80M & 1.95G & -11dBm & 1233\\
% \hline
% WiFi & 80M & 1.95G & -12dBm & 1012\\
% \hline
% WiFi & 80M & 1.95G & -13dBm & 1000\\
% \hline
% WiFi & 80M & 1.95G & -5dBm & 1004\\
% \hline
% WiFi & 80M & 1.95G & -0dBm & 952\\
% \hline
% WiFi & 80M & 3.49G & -11dBm & 1095\\
% \hline
% WiFi & 80M & 3.49G & -12dBm & 1068\\
% \hline
% WiFi & 80M & 3.49G & -13dBm & 1233\\
% \hline
% \end{tabular}
% \label{waveform}
% \end{center}
% \end{table}

% \begin{table*}[h]
% \caption{Jamming Waveforms Description}
% \begin{center}
% \begin{tabular}{|c|c|c|c|}
% \hline
% \textbf{Protocol Type} & \textbf{Bandwidth (MHz)}& \textbf{Center Frequency (MHz)}& \textbf{Power Levels (dBm)} \\
% \hline
% WiFi & 80  & 2140  & 0, -5, -11, -12, -13  \\
% \hline
% WiFi & 80  & 1950  & 0, -5, -11, -12, -13\\
% \hline
% WiFi & 80  & 3490  & -11, -12, -13 \\
% \hline
% \end{tabular}
% \label{waveform}
% \end{center}
% \end{table*}

% Please add the following required packages to your document preamble:
% \usepackage{booktabs}
\begin{table*}[]
\centering
\caption{Attack Dataset Table with Description of Jamming Waveforms}
\label{tab:waveforms_description}
\begin{tabular}{@{}cccc@{}}
\toprule
\textbf{Attack Waveform Protocol} & \textbf{Bandwidth (MHz)} & \multicolumn{1}{l}{\textbf{Center Frequency (MHz)}} & \multicolumn{1}{l}{\textbf{Transmit Power Levels (dBm)}} \\ \midrule
802.11n WiFi & 80 & 1950 (LTE cell uplink band) & 0, -5, -11, -12, -13  \\ \midrule
802.11n WiFi & 80 & 2140 (LTE cell downlink band) & 0 , -5, -11, -12, -13 \\ \midrule
802.11n WiFi & 80 & 3490 (5G NR cell band) & -11, -12, -13         \\ \bottomrule
\end{tabular}
\end{table*}
As depicted in Fig.~\ref{fig:system_model}, the legitimate communications link between the BS and the UE is 5G NSA with i) a New Radio (NR) cell Time Division Duplex (TDD) in band $N78$ and ii) an LTE cell Frequency Division Duplex (FDD) in band $B1$. The Center Frequency (CF) for the jamming signal is selected as the respective CF for band $N78$ ($3.50$ GHz), uplink band $B1$ ($1.95$ GHz), and downlink band $B1$ ($2.14$ GHz). Different power levels of the jamming signal are selected to compare their respective impact on the performance metrics measured at the UE. At these power levels, the legitimate communications link is still affected but without a call drop which is the area of interest for this work. 
In this implementation, the duration between consecutive time samples is $180$ ms, i.e., it is the period at which consecutive samples of $\mathbf{Y}[n]$ are logged.
% From Table~\ref{waveform}, the maximum jamming power level at the NR CF is less than that in the LTE CF because, with equivalent power levels of jamming, NR cells are more vulnerable to call drops. 

Two broad categories of supervised machine learning models were constructed and evaluated for comprehensive coverage. The first category is an instantaneous discriminative model which is used for detecting known types of interference within short intervals ($<180$ ms). The second type is a time series-based sequential model, typically based on neural networks, e.g., those from the recurrent neural network (RNN) family. Such sequential models are better suited for detecting known types of interference having a relatively longer duration ($>180$ ms), and therefore, can exploit underlying temporal correlations. 
% An unsupervised learning-based anomaly detection model is used for detecting unknown types of jamming. The computational complexity and required number of data samples for training and evaluation of these three types of models increase incrementally. 
Sec.~\ref{sec:machinelearningmodel} provides the detailed architecture of these two categories of models and their corresponding performance evaluation using the dataset generated as described in Table~\ref{tab:waveforms_description} is detailed in Sec.~\ref{sec:performance_evaluation}. 
% and accuracy of the models are expected to increase, especially the latter two as the sample size increases. 

\subsection{Supervised Learning-based Detection Models}\label{sec:machinelearningmodel}

The designed jamming detection module 
% for known jamming types \textcolor{blue}{(i.e., assuming knowledge of protocol of the jamming waveform)} 
feeds $\mathbf{Y}(n)$ in Eq.~\eqref{eq:temporal_data} as the input into a discriminative predictor. 
An instantaneous classification model based on a single time step of data ($T=0$ in Eq.~\eqref{eq:temporal_data})  needs shorter observation time intervals, whereas a time series-based sequential model (typically recurrent neural networks or an attention-based model, e.g., transformer) requires a sequence of data across consecutive time steps ($T>0$ in Eq.~\eqref{eq:temporal_data}), thereby revealing and exploiting underlying temporal correlation in its classification decision.
% A temporal-based model can preserve the frequency characteristics of the wireless signal, and a single time stamp classification takes less observation time in the field application. 
We consider and evaluate the performance of both instantaneous and sequential learning models in this work. 
% Instantaneous classification is built with a single time stamp of data, which has the advantage of fast detection time within a single sample period. 
% In our implementation, the duration between consecutive time samples is $180$ ms, i.e., the sampling period. 
% Sequential learning is based on $T$ consecutive time steps, and thus provides $T$ consecutive data samples as the input sequence to the reccurent neural network model for detection. 
In the training and validation stage, the type of jamming information shown in Table~\ref{tab:waveforms_description} is automatically recorded and labeled in a local database. The $N_f = 34$ parameters collected ($17$ for each cell ID, i.e., $17$ features from the NR cell and the LTE cell) for each time step, which are subsequently used as input features to both the instantaneous model and the sequential model(s), are the following: i) Cell ID, ii) Downlink and Uplink bit rates,
% downlink/uplink bitrate,
iii) Downlink and Uplink packet rates, iv) Downlink and Uplink Re-transmission rates, 
% downlink/uplink packet rate,
v) PUSCH SNR, vi) Channel Quality Indicator (CQI), vii) Power Headroom, viii) Energy per Resource Element (EPRE), ix) Uplink Path Loss, x) Downlink and Uplink Modulation and Coding Scheme (MCS) values, and xi) Minimum, Average and Maximum turbo decoder rates.
% \textcolor{red}{Where is the turbo decoder running? UE or the BS?}
The above independent variables used as input features to ML models are selected based on the criteria of not impacting legitimate communication quality, which sets the foundations of real-world deployments and applications.

\subsubsection{Instantaneous Classification Models}
% The designed jamming detection module for known jamming types feeds $y(n)$ in Eq.~\eqref{model input} as the input into a discriminative predictor. A temporal-based model can preserve the frequency characteristics of the wireless signal, and a single time stamp classification takes less observation time in the field application.  We have considered the performance for both instantaneous classification and temporal-based learning.
% Instantaneous classification is built with a single time stamp of data, which has the advantage of fast detection time within one sample period. In our implementation, the duration between consecutive time samples is 180 ms, i.e., the sampling period. In the training and validation stage, the type of jamming information shown in Table~\ref{waveform} is automatically recorded and labeled in a local database. The independent variables used as input features to both the instantaneous classification model as well as the temporal-based detector are downlink/uplink bitrate, downlink/uplink bitrate, downlink/uplink packet rate, downlink/uplink retransmission rate, downlink/uplink packet rate, PUSCH SNR, CQI, power headroom, energy per resource element, uplink path loss, downlink/uplink MCS, and turbo decoder rate. 
% The discriminative models are trained using positive (known target jamming types) and negative cases (no jamming present).
% \textcolor{blue}{(Add an introductory statement).}
The instantaneous jamming detection task is a discriminative in nature where, given a set of $N_{\text{tr}}$ feature vectors $\{\mathbf{r}_i \in \mathbb{R}^{N_f} \}_{i=1}^{N_{\text{tr}}}$ for training and $\mathcal{C}$ representing the space of all classifiers belonging to a particular family of instantaneous classifiers, the learning algorithm ``learns'' a mapping $c \in \mathcal{C}$ given by $c(\mathbf{r}_i): \mathbb{R}^{N_f} \rightarrow \{0,1 \}$.
% where the machine learning model learns a mapping: $\mathbf{h}\in \mathbb{R}^{N_f}$ \textcolor{blue}{(Add mapping equation from Sayed book).} 
Thus, we consider a collection of classifier model families that take as input a single time-step of data, i.e., a single feature vector and provide a prediction for the presence or absence of jamming, i.e., a binary discriminative result. The classification models considered include: Logistic Regression, $K$-Nearest Neighbors, Gaussian Naive Bayes classifier, Random Forest classifier, AdaBoost classifier and Bagging classifier in particular. 
The primary advantage of these instantaneous discriminative models is the extremely low training and inference (testing) computational complexities, resulting in corresponding low train and inference run times, which are critical for deployment in low-power IoT devices or smartphones (UEs), where power consumption and adhering to latency constraints is critical.
However, this comes at the cost of considerably poorer overall detection performance.
This trade-off is made more extreme when considering multi-scenario classification, as we shall see in the performance results of Sec.~\ref{sec:multi_scenario_classification_performance_evaluation}.
The details of their implementation and a detailed performance evaluation is provided in Sec.~\ref{sec:performance_evaluation}. 

\subsubsection{Sequential Time-Series based Models}
% \textcolor{blue}{(First, describe the multivariate time-series classification problem. Lay the foundation for a generic RNN structure, then motivate all the other structures with this notation.)}
A standard recurrent neural network (RNN) can be used for a multivariate time series classification.
With this setup, the input to the model is now a ``sequence'' of length $T$ denoted as $\mathbf{Y}_i \in \mathbb{R}^{N_f \times T}$. Thus, given a set of $N_{\text{tr}}$ such sequences $\{ \mathbf{Y}_i \}_{i=1}^{N_{\text{tr}}}$ for training and for a class $\mathcal{C}_r$ of a particular family of RNN models, the learning algorithm learns a mapping $c \in \mathcal{C}_r$ given by $c(\mathbf{Y}_i): \mathbb{R}^{N_f \times T} \rightarrow \{0,1\}$.
Conventional or ``vanilla'' RNNs are known to be challenging to train using BPTT, specifically due to the problem of exploding and vanishing gradients~\cite{pascanu2013difficulty}. 
Primarily, two alternate recurrent structures, namely the LSTM and the GRU have been introduced to combat this issue, which we overview briefly in the following and will incorporate in our detector and classifier models.

\textbf{Long Short-Term Memory (LSTM):}
The LSTM (Long Short-Term Memory)~\cite{Hochreiter1997} is a variant of the RNN which by incorporates three gating mechanisms, known as the input gate, forget gate and output gate. This gates function as valves, controlling the amount of information that enters into the LSTM units, that gets stored its internal memory and that exits the LSTM unit.
Due to this architecture and owing to the higher number of trainable parameters compared to RNNs, LSTMs have richer modeling capabilities and provide more robust performance compared to RNNs while also alleviating the problem of vanishing and exploding gradients. However, this comes at the cost of additional training complexity requiring training LSTMs for higher number of training epochs.
% common variant of an RNN
% The LSTM has a chain structure consisting of a specific neural network cell structure. 
% The cell consists of three gates: input gate, output gate, and forget gate. 
% Traditional RNNs and their variants such as the LSTM are trained using gradient-based algorithms such as Back Propagation Through Time (BPTT).
% While the LSTM avoids the vanishing and exploding gradients problem, it has a considerable training computational complexity and requires a large number of data samples to achieve a desired training loss. 
% \textcolor{blue}{(new start)}
% An illustrative block diagram showing the internal details of a single LSTM unit is shown in Fig. XXX.
% \textcolor{blue}{(new end)}

\textbf{Gated Recurrent Unit (GRU):}
The Gated Recurrent Unit (GRU)~\cite{cho-etal-2014-learning} implements an alternate gating mechanism compared to the LSTM, in that it has a forget gate and a remember gate compared to the three gates in the LSTM unit. While the two gates also modulate information flow through the gate, their training complexity is lower compared to the LSTM, and therefore can achieve comparable performance in most tasks. 
% employs a similar gated architecture as the LSTM in order to modulate the amount of information flow through the GRU unit.
% However, it uses fewer parameters than the LSTM
% An improvement over LSTMs is suggested in the form of the Gated Recurrent Unit (GRU)~\cite{cho-etal-2014-learning}.
% The GRU accomplishes the same result but with fewer parameters than the LSTM via an alternate gating mechanism.

\begin{figure}[!h]
    \centering    \includegraphics[width=\linewidth]{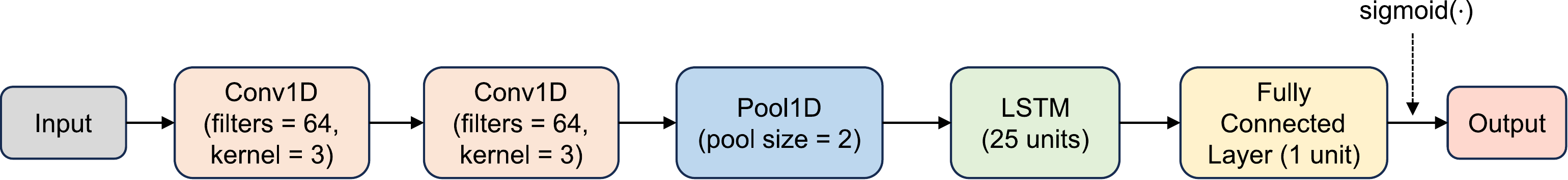}
    \caption{Architecture of the CNN-LSTM (appended with convolutional layers and pooling layers at the input of the LSTM units) for binary jamming detection.}
    \label{fig:cnn_lstm_arch_binary_detection}
\end{figure}

\textbf{Convolutional Input-based LSTM:}
% \subsubsection{Convolutional Input based LSTM}
An improvement over using a simple recurrent unit such as the LSTM or the GRU for sequence classification can be to use convolutional layers in order to extract features from the raw input before feeding the extracted feature sequence to the recurrent structure for further classification, as originally proposed in~\cite{Sainath2015}. We qualify such an architecture as the `CNN-LSTM' model (originally named `CLDNN' in~\cite{Sainath2015}) and show the detailed architecture of our implementation of the model for jamming detection in Fig.~\ref{fig:cnn_lstm_arch_binary_detection}. 
% \textcolor{blue}{Add figure number}

\begin{figure}[!h]
    \centering    \includegraphics[width=\linewidth]{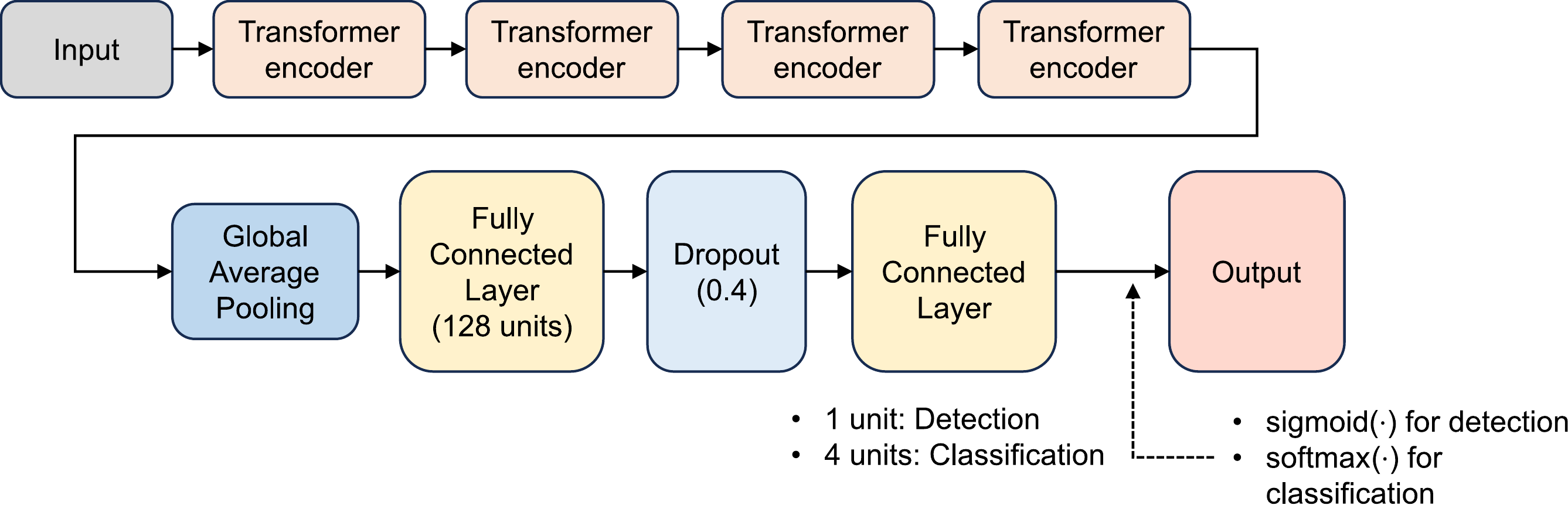}
    \caption{Architecture of the transformer model~\cite{Vaswani2017} adapted for discriminative output (detection or classification).}
    \label{fig:transformer_arch}
\end{figure}

\textbf{Self Attention-based Transformer:}
% \subsubsection{Self-Attention based Transformer}
The attention mechanism~\cite{kim2017structured, parikh-etal-2016-decomposable} has been widely employed in the field of natural language processing for a variety of sequence modeling tasks.
The transformer architecture introduced in~\cite{Vaswani2017} relies completely on the self-attention mechanism and does not invoke recurrence or convolutions, thereby enabling parallelization.
The transformer encoder comprises a multi-head self-attention (MHA) layer connected to fed to a feedforward neural network layer. Both layers employ residual or `skip' connections.
We adapt the original transformer architecture for sequence classification as opposed to sequence to sequence (seq2seq) conversion by replacing the decoder layer of the original model with pooling followed fully connected layers to give discriminative classification decision outputs. 
The block diagram of our implementation is given in Fig.~\ref{fig:transformer_arch}.
% A complete block diagram for this architecture can be found in~\cite{Vaswani2017}. 
% The multi-head self-attention (MHA) layer is the basic structure inside the encoder block of the transformer architecture, the output of which is passed to a feedforward neural net
% along with positional encoders and decoders that are fulfilled with feedforward neural networks.
% \textcolor{blue}{The parameters used in our implementation of the Transformer are provided in Table XXX}. 
% A visual description of the self-attention based transformer model is shown in Fig. ZZZ.\textcolor{blue}{Add figure number} 

\begin{figure}[h]
    \centering
    \includegraphics[width=0.85\linewidth]{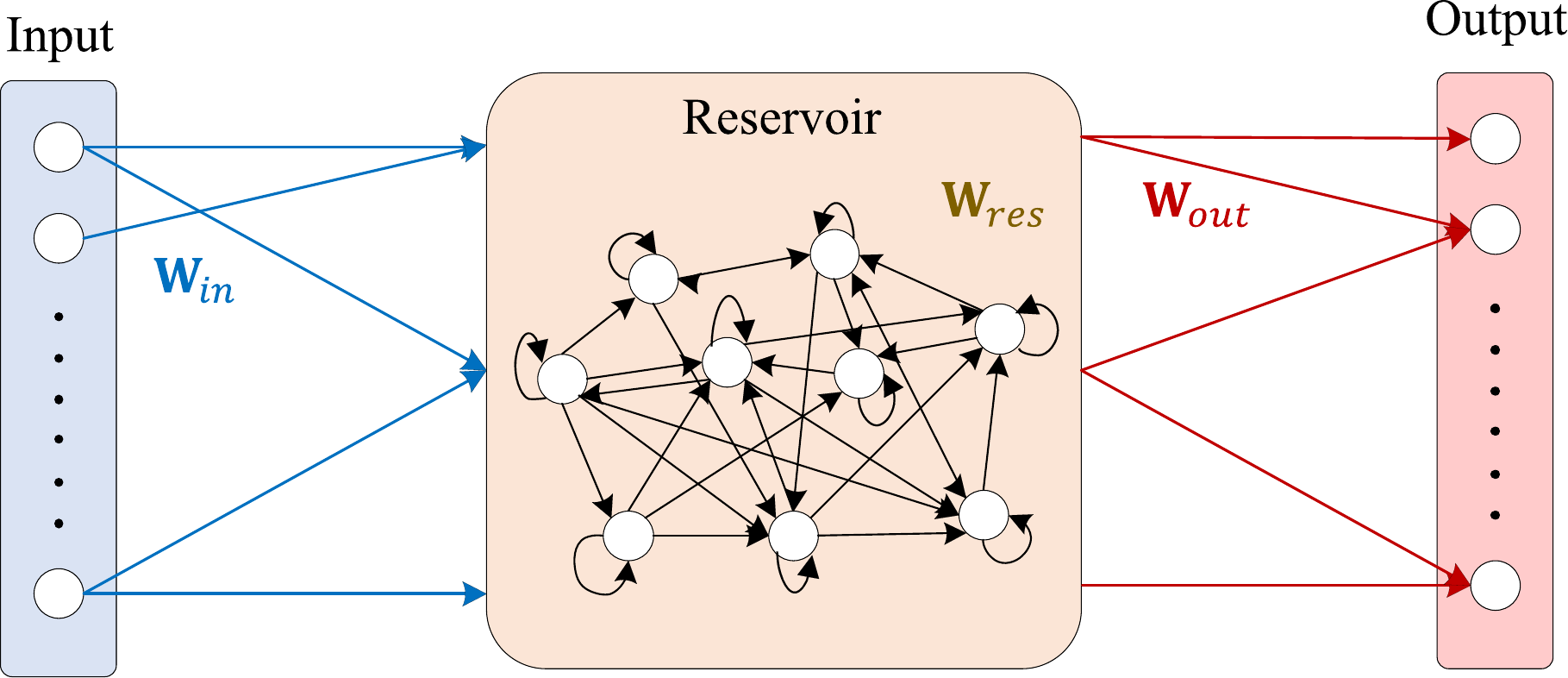}
    \caption{A single-reservoir Echo State Network (ESN).}
    \label{fig:ESN_figure}
\end{figure}

\textbf{Reservoir Computing (RC)-based model:}
% \subsubsection{Reservoir Computing (RC)-based model}
While architectural improvements offered by LSTM and GRU alleviate the vanishing gradient problem to some extent, the Reservoir Computing (RC) framework within the broader category of randomized recurrent neural networks~\cite{Gallicchio2018RandomizedRN} offers an alternate paradigm and avoids these problems encountered in training conventional RNNs. Specifically, we utilize the echo state network (ESN)~\cite{lukovsevivcius2009reservoir}, which is a particular implementation of the RC paradigm and requires training a significantly lower number of network (model) weights as compared to vanilla RNNs. 
% More information about the ESN and its applications especially in wireless communications tasks can be found in []. 
An ESN with a single ``reservoir'' along with its relevant parameter weights matrices is depicted in Fig.~\ref{fig:ESN_figure}. In the following, we define the relevant input, output, reservoir and feedback weights matrices for the ESN structure. 
\begin{itemize}
% \vspace{-2.5mm}    
    \item $\mathbf{x}_{\text{in}}(t) \in \mathbb{R}^{N_f}$ denotes the input at discrete time index $t$, where $N_f$ is the input dimension (i.e., number of input features). $\mathbf{x}_{\text{res}}(t) \in \mathbb{R}^{M}$ denotes the state vector at the discrete time index $t$ within the input sequence of length $T$, with $M$ being the number of neurons in the reservoir. 
    % Define $\mathbf{X}_{\text{res}} = \left[\mathbf{x}_{\text{res}}(1), \cdots, \mathbf{x}_{\text{res}}(T) \right] \in \mathbb{R}^{M \times T}$ as the ``reservoir states matrix'' of the individual states across time from $t=1$ to the end of the input sequence $t=T$ stacked sequentially.
    % \vspace{-1.5mm}
    \item $\mathbf{W}_{\text{in}} \in \mathbb{R}^{M \times N_f}$ denotes the input weights matrix, $\mathbf{W}_{\text{res}} \in \mathbb{R}^{M \times M}$ denotes the reservoir weights matrix and $\mathbf{W}_{\text{out}} \in \mathbb{R}^{K \times M}$ denotes the output weights matrix. 
    % and $\mathbf{W}_{\text{fb}} \in \mathbb{R}^{M \times K}$ is the feedback weights matrix when teacher forcing~\cite{Lukosevicius2012} is enabled. 
    $\mathbf{y}(t) \in \mathbb{R}^{K}$ denotes the ESN output.
    % $\mathbf{W}_{in}^{(i)} \in \mathbb{R}^{M \times D}$ for $i=1$ and $\mathbf{W}_{in}^{(i)} \in \mathbb{R}^{M \times K}$ for $i=2,\cdots,L$.
    % \vspace{-1.5mm}
    % \item $\mathbf{W}_{\text{res}} \in \mathbb{C}^{M \times M}$ is the reservoir weights matrix.
    % \vspace{-1.5mm}
    % \item $\mathbf{W}_{\text{out}} \in \mathbb{C}^{K \times M}$ is the output weights matrix.
    % \vspace{-1.5mm}
    % \item $\mathbf{W}_{\text{fb}} \in \mathbb{C}^{M \times K}$ is the feedback weights matrix when teacher forcing~\cite{Lukosevicius2012} is enabled.
    % \vspace{-1.5mm}
    % \item $\mathbf{y}(t) \in \mathbb{C}^{K}$ denotes the ESN output. 
    % Define $\mathbf{Y}^{(i)} = \left[\mathbf{y}^{(i)}(1), \cdots, \mathbf{y}^{(i)}(T) \right] \in \mathbb{R}^{K \times T}$ as the matrix of outputs stacked together.
\end{itemize}

% Let $t = \{1,2,\cdots,T\}$ denote the discrete-time indices in a particular input sequence $\mathbf{U}$. 
% For a single-reservoir ESN structure, its input $\mathbf{x}_{\text{in}}(t)$ is simply $\mathbf{u}(t)$.
If $\sigma(\cdot)$ is a pointwise non-linear activation function, e.g., ReLU or $\tanh(\cdot)$ and $\sigma_{o}(\cdot)$ is the sigmoid or the softmax activation function (for classification tasks), then the state update equation and the output equation are respectively:
\begin{align}
\mathbf{x}_{\text{res}}(t) &= \sigma \big(\mathbf{W}_{\text{res}}\mathbf{x}_{\text{res}}(t-1) + \mathbf{W}_{\text{in}}\mathbf{x}_{\text{in}}(t)  \big) \nonumber \\
&+ (1-\alpha)\mathbf{x}_{\text{res}}(t-1), \\
%\end{align}
%\vspace{-2.5mm}
%\begin{align}
\mathbf{y}(t) &= \sigma_{o}\big( \mathbf{W}_{\text{out}}\mathbf{x}_{\text{res}}(t)\big),
\label{eq:output_eqn}
\end{align}
where $\alpha \in [0,1]$ is the reservoir leakage rate.
Note that for the binary jamming detection task, $K=1$ and $\sigma_{o}(\cdot)$ is the sigmoid activation function while for the multi-scenario classification task, $K=N_{\mathrm{jam}}=4$ (in this work) and $\sigma_{o}(\cdot)$ is the softmax activation function.
% In Fig.~\ref{fig:ESN_figure}, $\mathbf{W}_{\text{in}}$ is the input weights matrix, $\mathbf{W}_{\text{res}}$ is the reservoir weights matrix and $\mathbf{W}_{\text{out}}$ is the output weights matrix. 
In the ESN structure, $\mathbf{W}_{\text{out}}$ is the only trainable parameter. Other parameters such as $\mathbf{W}_{\text{in}}$, $\mathbf{W}_{\text{res}}$ are initialized according to a pre-determined distribution (e.g., uniform distribution $\mathcal{U}(-1,1)$) and kept unchanged throughout training and inference. This reduces the number of trainable parameters in the RC framework significantly and lends itself to applications with limited availability of training data.
The hyperparameter $\rho$, known as the spectral radius is defined as $\rho = \max |\lambda_{\text{eig}}(\mathbf{W}_{\text{res}}) |$, where $\lambda_{\text{eig}}(\mathbf{W}_{\text{res}})$ denotes the set of eigenvalues of $\mathbf{W}_{\text{res}}$.
The hyperparameter $\kappa$, known as sparsity, is the fraction of total elements in $\mathbf{W}_{\text{res}}$ that are zero.
Unlike using a gradient-based algorithm such as BPTT, only the readout (output) layer needs to be trained, which can be performed by solving a simple Least Squares (LS) regression problem, resulting in a closed-form solution involving the matrix pseudo-inverse. 
It has been shown that in general for RNNs, bidirectional architectures can extract long-term dependencies in the input sequence that are widely spaced in time~\cite{GRAVES2005602}. 
This is true of reservoirs in RC as well~\cite{bianchi2018bidirectional}, especially for the classification task where the entire input sequence is available for bidirectionality to be exploited.
For a bidirectional reservoir setup, the dynamics equations are updated as follows:
\begin{align}
    \overrightarrow{\mathbf{x}}_{\text{res}}(t) &= \sigma \big(\mathbf{W}_{\text{res}}\overrightarrow{\mathbf{x}}_{\text{res}}(t-1) + \mathbf{W}_{\text{in}}\overrightarrow{\mathbf{x}}_{\text{in}}(t) \big), \nonumber \\
    &+ (1-\alpha)\overrightarrow{\mathbf{x}}_{\text{res}}(t-1), \nonumber \\    
    \overleftarrow{\mathbf{x}}_{\text{res}}(t) &= \sigma \big(\mathbf{W}_{\text{res}}\overleftarrow{\mathbf{x}}_{\text{res}}(t-1) + \mathbf{W}_{\text{in}}\overleftarrow{\mathbf{x}}_{\text{in}}(t) \big) \nonumber \\
    &+ (1-\alpha)\overleftarrow{\mathbf{x}}_{\text{res}}(t-1).
\end{align}
Here, $\overleftarrow{\mathbf{x}}_{\text{in}}(t) = \overrightarrow{\mathbf{x}}_{\text{in}}(T-t)$ and $\overrightarrow{\mathbf{x}}_{\text{in}}(t) \in \mathbb{R}^{N_f}$ is the $t^{\text{th}}$ sample of the sequence in its original order. The complete state vector at any time instant $t$ is given by $\mathbf{x}_{\text{res}}(t) = [\overrightarrow{\mathbf{x}}_{\text{res}}(t); \overleftarrow{\mathbf{x}}_{\text{res}}(t)] \in \mathbb{R}^{2M}$.
Correspondingly, the weights matrices also increase in size according to $\mathbf{W}_{\text{in}}^{(b)} \in \mathbb{R}^{2M \times N_f}$, $\mathbf{W}_{\text{res}}^{(b)} \in \mathbb{R}^{2M \times 2M}$ and $\mathbf{W}_{\text{out}}^{(b)} \in \mathbb{R}^{K \times 2M}$.
For a total of $N_{\text{tr}}$ sequences in the training dataset, the labels for each of which are collected in a one-hot encoded class membership matrix (ground truth) $\mathbf{G} \in \mathbb{R}^{K \times N_{\text{tr}}}$, define the reservoir states matrix $\mathbf{X}_{\text{res}} \in \mathbb{R}^{2M \times N_{\text{tr}}}$ which collects the final state vector $\mathbf{x}_{\text{res}}(T)$ in each training input sequence. Then the trainable weights $\mathbf{W}_{\text{out}}^{(b)}$ are found by solving the following regularized least-squares optimization:
\begin{align}
    \widehat{\mathbf{W}}_{\text{out}}^{(b)} = \argmin_{\mathbf{W}_{\text{out}}^{(b)}} \left( \|\mathbf{G} - \mathbf{W}_{\text{out}}^{(b)} \mathbf{X}_{\text{res}} \|_{2}^{2} + \lambda \|\mathbf{W}_{\text{out}}^{(b)} \|_{2}^{2} \right),
\end{align}
where $\lambda$ is the regularization constant. This allows for a simple pseudo-inverse based update rule that has a significantly lower complexity compared to iterative gradient descent based backpropagation algorithms employed to train conventional recurrent neural network architectures.
% The hyperparameters used in our implementation of the bidirectional ESN are outlined in \textcolor{blue}{Table ZZZ}.

The advantages of employing an ESN have been extensively studied in our previous work, e.g.,~\cite{Jere2023TCOM}, where we theoretically show the superior generalization ability of ESNs compared to standard RNNs using tools from statistical learning theory. 
Furthermore in~\cite{Jere2023WCL}, we have developed a classical signal processing understanding of the ESN and derived an analytical optimum for the conventionally untrained reservoir weights when applied for wireless channel equalization.
Additionally, in our most recent work~\cite{Jere2023universal}, we make the case for ESNs being universal approximators of linear time-invariant (LTI) systems and have derived an analytical expression for the optimal probability density function (PDF) of the reservoir weights which are otherwise randomly initialized in practice, e.g., from the uniform or Gaussian distribution, when the ESN is employed for LTI system identification.
Building upon this solid analytical footing of our prior theoretical work showing the effectiveness of ESNs, we incorporate the concept of bidirectionality which has empirically demonstrated performance benefits~\cite{GRAVES2005602, bianchi2018bidirectional}.

\subsection{Bayesian Causality Analysis}\label{sec:causal}
% Suppose a jamming or interference type $J_{i}$ coexists with a legitimate user, and the Key Performance Indicator (KPI) pattern $P_{k}$ is observed with the received signal as the outcome of the jamming or interference causal effects. Then,
Assuming the presence of a jamming or interference entity $J_{i}$ alongside a legitimate user, we observe the Key Performance Indicator (KPI) pattern $P_{k}$ in the received signal as a result of the causal effects of the jamming or interference. Then, we define the following hypotheses:
\begin{itemize}
    \item $Y^{i}_{k} = 1$: Pattern $P_{k}$ is detected if the jamming or interference type is $J_{i}$.

    \item $Y^{i}_{k} = 0$: Pattern $P_{k}$ is detected if the jamming or interference type is not $J_{i}$.
\end{itemize}

% One of the challenges of detecting the presence of jamming/inference through a specific KPI pattern detection is that the same pattern that is detected could be caused by different direct, indirect, and root causes. Detecting the root causes could provide fundamental discoveries to jamming/interference prevention and system resilience. Additionally, domain knowledge and experience play an important role in detecting the causes of performance degradation and addressing security concerns. Thus, we have introduced a causation model in this study to root cause external factors focusing on jamming and inference. This model can be expanded to include general external or internal factors across multiple network layers and application types.
Detecting the presence of jamming or interference through a specific KPI pattern is challenging since the same pattern may result from various direct, indirect, or root causes. Identifying the root causes can lead to significant breakthroughs in mitigating the jamming and interference present and enhancing system resilience. Furthermore, domain knowledge and expertise are vital in detecting the causes of performance degradation and addressing security concerns. Therefore, our study introduces a causation model that focuses on identifying the external factors focusing on jamming and interference. This model can be expanded to encompass both external and internal factors across various network layers and application types.
The causation model employs the ``no causation without manipulation'' approach \cite{Holland1986StatisticsInference} to detect causation. This requires integrating domain knowledge with manipulation or intervention on mutable variables, such as different types and combinations of jamming or interference. By establishing the causal effects between these variables, indirect causes, and observed KPI patterns, certain actions can be taken to improve performance by reducing or eliminating their impact on the relevant KPIs. In this paper, we define causal effects as the average causal effects $E[Y^{i}]$. This is the difference between the average value of $Y$ when jamming or interference type $J_{i}$ is present and when it is not.
To facilitate causal effect analysis, we make two assumptions: consistency and ignorability. Consistency implies that the potential outcome of $Y$ is equal to the observed outcome $Y^i$ if jamming or interference type $J_i$ is present. Ignorability assumes that the choice of a cyber-attack (in the form of a jamming attack in this work) is not influenced by the current KPI of the network. We plan to investigate interactive or adaptive cyber-attacks through the Markov Random Field (MRF) in future work.
Causal inference models allow the combination of machine learning and expert knowledge into explainable machine learning (XML). The Bayesian Network Model (BNM) is one such approach in the causal inference arsenal. This technique allows for both quick and accurate results while incorporating expert input based on domain knowledge, leading to better models even with limited training dataset sizes. Querying a Bayesian Network provides immediate insight into the importance and influence of each variable on a specific outcome.

% In the BNM, immediate and direct expert knowledge is used to control the existence and direction of edges between graph nodes, thereby encoding knowledge into a Directed Acyclic Graph (DAG) with distributed probability. Therefore, the critical step in building the causal inference model via the BNM is the establishment of the DAG through the expert knowledge. In the application of BNM in anonymous jamming and interference, this step is performed by constructing the DAG and the probability distribution of its relationship between nodes. 
% Learning DAGs from data is an NP-hard problem\cite{Chickering2004Large-sampleNP-hard}, owing mainly to the combinatorial acyclicity constraint that is difficult to enforce efficiently. A heterogeneous domain rule-based DAG construction for jamming and interference detection is proposed in our study. The rules from domain knowledge are proposed and formatted to generate the initial structure of the DAG. Thereafter, data-driven auto-construction models are added to the DAGs for a more comprehensive view. 
In the BNM, domain-specific knowledge is utilized to determine the existence and direction of edges between graph nodes, forming a directed acyclic graph (DAG) with distributed probability. Therefore, establishing the DAG through expert knowledge is a critical step in constructing a causal inference model using the BNM. In our application of the BNM for anonymous jamming and interference, we construct the DAG and the probability distribution of its relationship between nodes.
However, learning DAGs from data is an NP-hard problem \cite{Chickering2004Large-sampleNP-hard} due to the difficulty in enforcing the combinatorial acyclicity constraint. To overcome this challenge, we propose a heterogeneous domain rule-based DAG construction procedure for jamming and interference detection. Rules derived from domain knowledge are utilized to generate the initial structure of the DAG, followed by the addition of data-driven auto-construction models to provide a more comprehensive view.

\subsubsection{Rule Generation for Initial Construction}
\begin{figure*}[!h]
    \centering    \includegraphics[width=0.7\linewidth]{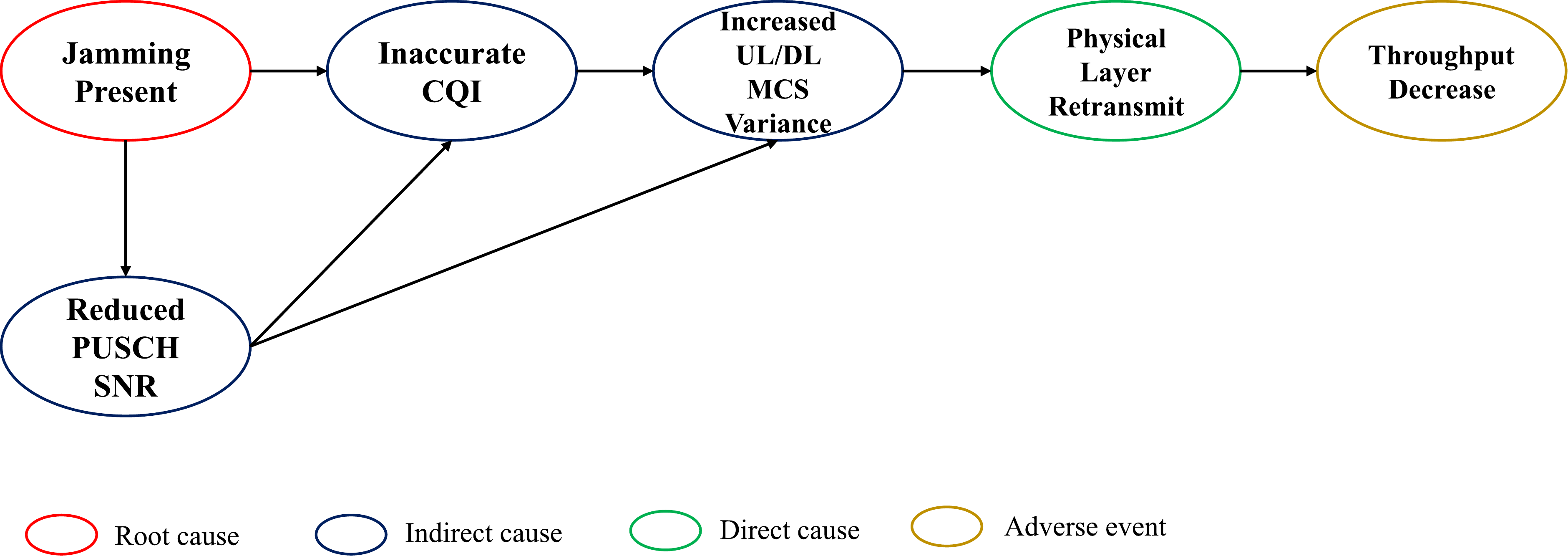}
    \caption{The chain of causal inference from the root cause to the observed adverse event, built using domain knowledge, e.g., relationships among observed KPI parameters from 3GPP standards.}
    \label{fig:causal}
\end{figure*}

% The PUSCH SNR (Signal-to-Noise Ratio) data collected from the testbed in decibel scale is first converted to the energy per bit, or $E_{b}/N_{0}$ metric. Then, the spectral efficiency $\eta$ is calculated according to $\eta = \log_{2}(1 + E_{b}/N_{0})$ for the purpose of looking up CQI values from existing tables referenced in specific 3GPP standard documents, i.e., Table 5.2.2.1-2 in~\cite{3GPP20205GSystem}. Based on the CQI index the equivalent modulation and code rate (MCS) is configured where the number of information bits is obtained from the Transport Block Size Determination calculations from Table 5.1.3.2 in~\cite{3GPP20205GSystem}. The logical relationship among SNR, Efficiency, CQI, MCS, throughput, and re-transmission rate is shown in an ontology graph shown in Fig.~\ref{fig:causal}. The ontology graph reveals the structure of BNM for causal inference based on domain knowledge. The underlining relationship of parameters and KPIs in domain knowledge and collected data can be further used to determine the probability distribution used for BNM. 
To begin, the PUSCH SNR data collected from the 5G NR testbed is converted from decibel scale to the energy per bit metric, denoted as $E_{b}/N_{0}$. Subsequently, the spectral efficiency $\eta$ is calculated using $\eta = \log_{2}(1 + E_{b}/N_{0})$ to lookup CQI values from existing tables referenced in specific 3GPP standard documents, such as Table 5.2.2.1-2 in~\cite{std3gpp38214}. Based on the CQI index, the corresponding modulation and coding scheme (MCS) index is configured, where the number of information bits is obtained from the Transport Block Size Determination calculations found in Table 5.1.3.2 in~\cite{std3gpp38214}.
The ontology graph shown in Fig.~\ref{fig:causal} displays the logical relationships among PUSCH SNR (and thus spectral efficiency), CQI, MCS, throughput, and re-transmission rate, revealing the structure of BNM for causal inference based on domain knowledge. This can be utilized to determine the probability distribution for the BNM, based on the underlying relationships between the parameters and the corresponding KPIs, gleaned from domain knowledge and the statistical distributions of the data collected from the testbed.

% \begin{figure}[!h]
%     \centering
%     \includegraphics[width=\linewidth]{figures/ontology_phy.png}
%     \caption{Ontology View for 5G Selected Parameters and KPI}
%     \label{fig:ontology_phy}
% \end{figure}

\subsubsection{Probability Distribution for BNM relationships}
After establishing the relationship among the nodes as seen in Fig.~\ref{fig:causal}, an essential component to complete the BNM is the probability distribution between the components. Following the rule of `no causation without manipulation', we calculated the with and without the conditional probability of each pair with directional causation on the BNM, generating an experiment-oriented probability distribution. We further use the specifications mandated by 5G NR standards to verify the observed probability and made adjustments if necessary. 
For the purpose of implementation and variable manipulation to compute posterior probabilities, the causal inference chain of Fig.~\ref{fig:causal} is implemented as a DAG in a software application such as Netica, with the generated BNM shown in Fig.~\ref{fig:bnm_2p1GHz_m13dBm}. 
The source node(s) is set as the condition, and the variable in the destination node is measured for its distribution under the condition of the source node(s). Given a new case for which we have limited knowledge, the developed BNM will find the appropriate values or probabilities for all the unknown variables for decision support. The ML module described in Sec.~\ref{sec:machinelearningmodel} can be viewed as a black-box, in which the available domain knowledge is extremely limited. 
With the domain knowledge-driven topology among different nodes and subsequently computing the posterior probabilities given the observed data, we can successfully rectify the incorrect detection decisions of the ML models with high confidence, as we demonstrate experimentally in Sec.~\ref{sec:bayesian_results}. 
% \textcolor{blue}{(Add a paragraph on why generative models cannot reliably be used for artificial training data generation).}

Although we use Bayesian causal inference in this work to mitigate potentially poor performance of the ML model caused by insufficient training data samples, another seemingly plausible alternative would be to use generative models for artificial training data generation such as Variational Autoencoders (VAEs) or Generative Adversarial Networks (GANs). However, such generative approaches can be prone to introducing bias caused by the small number of data samples used to train the generative model not being representative of the true (unknown) underlying joint distribution of the observed feature data and detection labels. Furthermore, GANs are known to be difficult to train, exhibiting several failure modes (e.g., mode collapse) and issues with convergence~\cite{Salimans2016}. Lastly, while generative methods can be applied when the jamming detection decisions are required within the infrastructure RAN or at the network edge, they may not be feasible for deployment on mobile UEs which have complexity and latency constraints.

% train multiple small models and integrate them into a Gray-box model which requires significantly less data to achieve the desired performance.  

% \subsection{Probability Distribution and Propagation}
% In this section, a probability distribution among the components in the ontology view and causal inference structure are derived from the data collected. The distribution between CQI and MCS shows an example in this process, and the probability distribution and propagation in other parts of the causal inference can be derived through the same method. 

% Independent features used in our model is shown in Table \ref{kpis}. 

% \begin{table}[htbp]
% \caption{Independent Variables for Antenna}
% \begin{center}
% \begin{tabular}{|c|c|c|}
% \hline
% cell\_id & dl\_bitrate & ul\_bitrate \\
% dl\_tx & ul\_tx & dl\_retx\\
% ul\_retx & pusch\_snr &epre \\
% cqi& ri& ul\_phr \\
% ul\_path\_loss& dl\_mcs& ul\_mcs \\
% turbo\_decoder\_min& turbo\_decoder\_avg&turbo\_decoder\_max\\
% ue\_index & max\_ue\_index&dl\_bitrate\_norm\\
% \hline
% \end{tabular}
% \label{kpis}
% \end{center}
% \vspace{-4mm}%Put here to reduce too much white space after your table 
% \end{table}

% \section{Jamming and Interference Segregation}
% \label{segregation}

\subsection{Supervised Learning-based Jamming Classification:}
\label{sec:jamming_classification_description}
% \textcolor{blue}{(Newly added subsection)}
In addition to deploying ML models for simple binary jamming detection, we also develop models for jamming scenario classification. Specifically, we focus on the ML models being able to classify the frequency band in which the jamming signal is present, as the knowledge of which control channel is being jammed can potentially aid in deploying combative strategies. 
Accordingly, we consider the following jamming scenarios for the ML model to classify: i) 1.95 GHz at 0 dBm, ii) 2.14 GHz at 0 dBm, iii) 3.49 GHz at -11 dBm and iv) No jamming signal present.
For this classification problem, we only consider employing the sequential models that take as input a sequence of consecutive time steps of data. This is because instantaneous models such the random forest classifier or the Naive Bayes classifier do not possess sufficient model capacity to discern patterns in temporally correlated data to perform $4$-class discriminative decisions based on only a single time step of feature data.
% Similar to the jamming detection task, we implement two categories of classifiers: i) Instantaneous and ii) Sequential or time series-based.
% \textcolor{blue}{(Finish this section).}

% \subsubsection{Instantaneous Classifier}
% For the multi-scenario classification task, we only consider the random forest model in this category. Other instantaneous models including the AdaBoost and Bagging classifiers display much worse performance with the limited training data available and are therefore not considered. Particularly for multi-scenario classification, the advantage of using time series-based sequential models that utilize multiple sequences of the temporally correlated input feature data is even more pronounced, thereby allowing the model to learn long-term dependencies and make better classification decisions.

\begin{figure}[!h]
    \centering
    \subfloat[GRU for classification.]{\includegraphics[width=\linewidth]{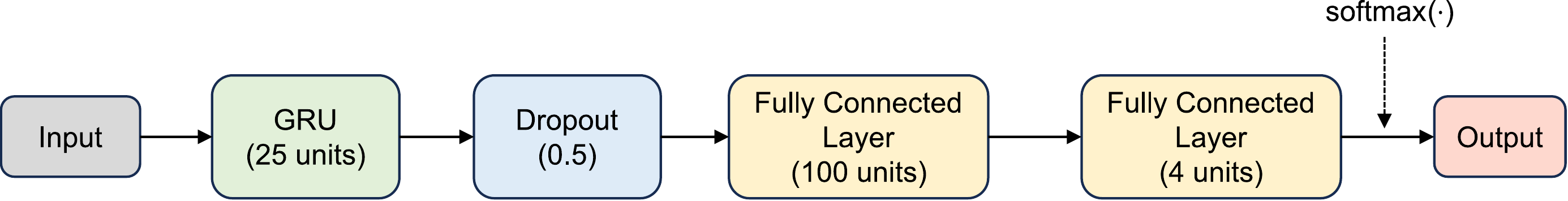}\label{fig:gru_arch_multi_scenario}}
    % \begin{subfigure}[h]{0.375\linewidth}
    %     \centering
    %     \includegraphics[width=\linewidth]{Maj_Rev_Figs/GRU_architecture_multi-scenario_classification.eps}
    % \caption{GRU for classification.}
    % \label{fig:gru_arch_multi_scenario}
    % \end{subfigure}    
    \vfill
    % \begin{subfigure}[h]{0.375\linewidth}
    %     \centering
    %     \includegraphics[width=\linewidth]{Maj_Rev_Figs/CNN-GRU_architecture_multi-scenario_classification.eps}
    % \caption{CNN-GRU for classification.}
    % \label{fig:cnn-gru_arch_multi_scenario}
    % \end{subfigure}
    \subfloat[CNN-GRU for classification.]{\includegraphics[width=\linewidth]{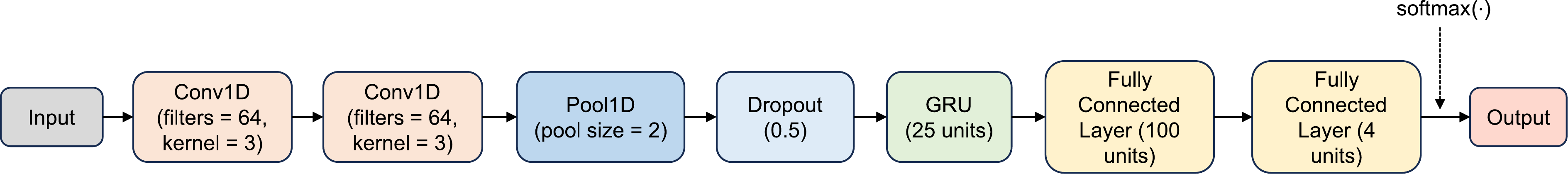}\label{fig:cnn-gru_arch_multi_scenario}}
    \caption{Architectures of the GRU-based (top) and the CNN-GRU (bottom) models (appended with convolutional and pooling layers at the input of the GRU units) employed for multi-scenario classification.}
\end{figure}

% \subsubsection{Time Series-based Classifiers}
The sequential models used for classification are similar to those employed for binary jamming detection, albeit with minor changes in architecture. We consider the GRU-based classifier with its detailed architecture shown in Fig.~\ref{fig:gru_arch_multi_scenario}. The CNN-GRU model, similar to the CNN-LSTM model used for detection is also considered for classification and shown in Fig.~\ref{fig:cnn-gru_arch_multi_scenario}. Finally, we consider the transformer model and the bidirectional RC (ESN) model which were also investigated for jamming detection. 
The performance evaluation for all of the above classification models is detailed in Sec.~\ref{sec:multi_scenario_classification_performance_evaluation} with the main conclusion that the RC-based approach can provide classification accuracy comparable to larger models but with significantly lower complexity.

% \begin{figure}[htbp]
%     \centering    \includegraphics[width=0.3\linewidth]{Maj_Rev_Figs/GRU_architecture_multi-scenario_classification.eps}
%     \caption{Architecture of the GRU-based recurrent model employed for multi-scenario classification.}
%     \label{fig:gru_arch_multi_scenario}
% \end{figure}

% \begin{figure}[htbp]
%     \centering    \includegraphics[width=0.3\linewidth]{Maj_Rev_Figs/CNN-GRU_architecture_multi-scenario_classification.eps}
%     \caption{Architecture of the CNN-GRU architecture (appended with convolutional and pooling layers at the input of the GRU units) employed for multi-scenario classification.}
%     \label{fig:cnn-gru_arch_multi_scenario}
% \end{figure}

\section{Performance Evaluation}
\label{sec:performance_evaluation}

\subsection{Binary Jamming Detection Performance}
\label{sec:performance_eval}
\begin{figure*}[h]
\centering
\includegraphics[width=0.95\textwidth]{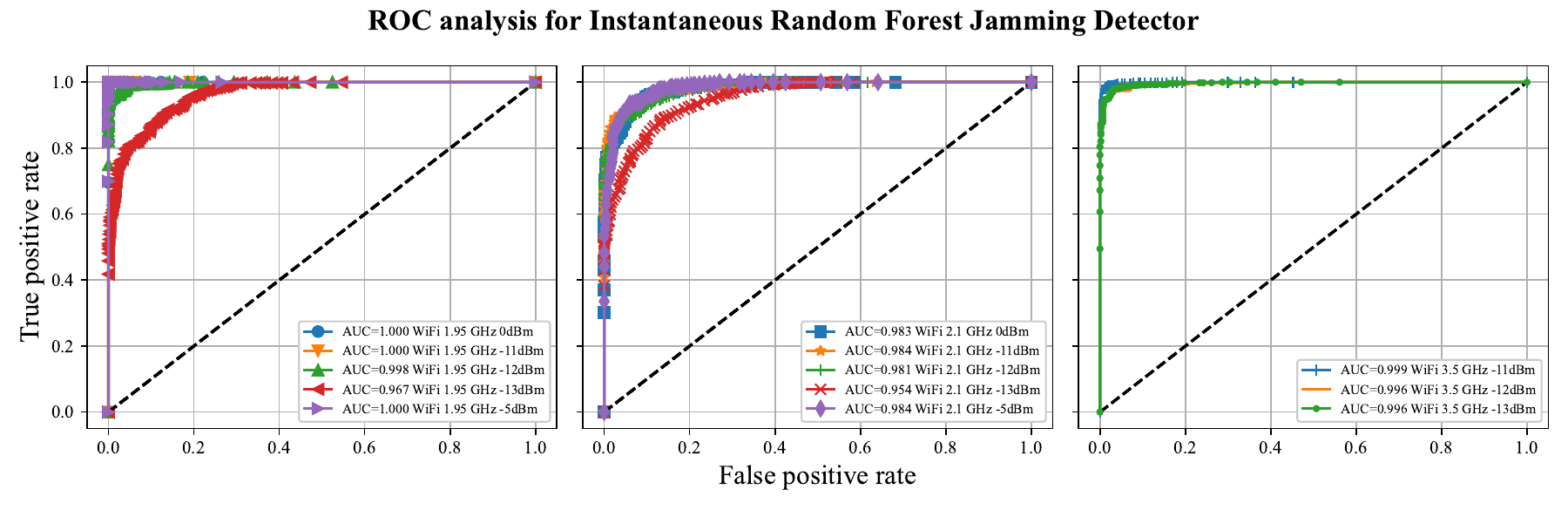}
\caption{Receiver Operating Characteristic (ROC) curves for each jamming scenario with the Instantaneous Random Forest detector.}
\label{roc_everything}
\end{figure*}

The performance evaluation of the introduced machine learning-based jamming detection models is based on the data collected as per the configurations of Table~\ref{tab:waveforms_description}. 
For instantaneous classification, a number of different models are used to compare the performance, namely: i) Random Forest, ii) AdaBoost, iii) Logistic Regression, iv) $K$-Nearest Neighbors, v) Gaussian Naive Bayes and vi) Bagging-based classifier. 
The jamming detection accuracy across the different scenarios is significantly high with ensemble learning motivated decision tree-based classification models. 
% When detecting the listed jamming types from the no jamming scenarios, the random forest algorithm gives $100\%$ accuracy for all jamming types. 
The reason behind this superior performance using decision trees is that decision trees or random forest models divide the feature space into much smaller regions, whereas other methods such as logistic regression fit a single line to divide the feature space exactly into two.
% The computationally lightweight tree-based classifier enables the possibility of migrating the interference detection system to a real-time implementation on IoT devices with low-power requirements. 
The Receiver Operating Characteristic (ROC) curves across all jamming scenarios along with the corresponding Area Under Curve (AUC) values for the Random Forest classifier is shown in Fig.~\ref{roc_everything}. The ROC curves are separated according to the band in which the interference is present, i.e., LTE UL ($1.95$ GHz), LTE DL ($2.1$ GHz) and 5G NR ($3.5$ GHz). The data split for training the model versus test is set to $75\%-25\%$ with $5$-fold cross-validation applied. 
Minmax normalization of all feature data is applied before being input for training or test to the instantaneous as well as sequential models.
We can see from Fig.~\ref{roc_everything} that across all scenarios, the lowest overall AUC is observed in the LTE DL band ($2.1$ GHz) with an interference power level of $-13$ dBm. This is in line with the expectation of the jamming signal with the lowest or close to the lowest detected power level having the lowest or close to the lowest detection accuracy.
An exhaustive performance evaluation of each instantaneous model is provided in Table~\ref{tab:jamming_detection_performance} which includes additional metrics such as Detection Rate (Accuracy), Precision, Recall, F-Score and the False Alarm Rate (FAR).
% In jamming classification for differentiating types of jamming, tree-based classification shows slightly lower but still significantly high performance. As an overview of the jamming classification methods, the AUC value to detect any specific jamming type against all other scenarios are shown in Fig.~\ref{roc_everything}

% \begin{figure}[htbp]
% \centering
% \includegraphics[width=0.6\textwidth]{figures/auc_heatmap.png}
% \caption{The Overview of AUC for Each Jamming Type}
% \label{auc_heatmap}
% \end{figure}

\begin{figure*}[h]
\centering
% \captionsetup{justification=centering}
% \includegraphics[width=0.8\textwidth]{Maj_Rev_Figs/roc_lstm_MajRev_2.eps}
\includegraphics[width=0.95\textwidth]{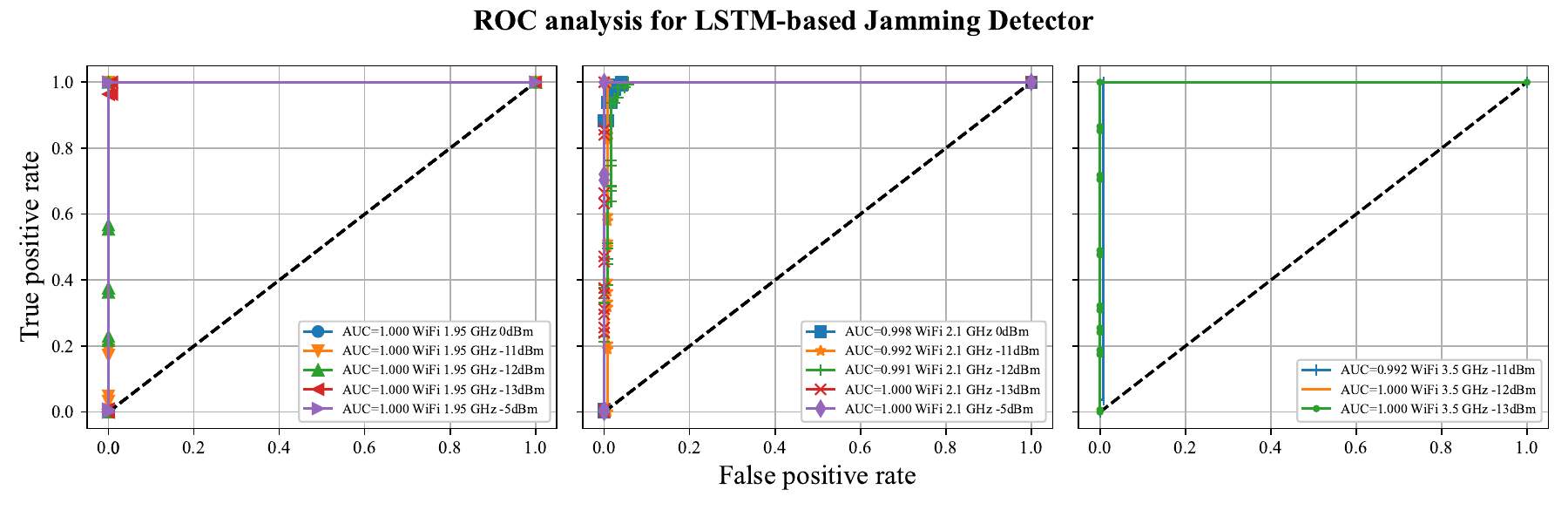}
\caption{Receiver Operating Characteristic (ROC) curves for each jamming scenario with the sequential LSTM-based detector.}
\label{roc_lstm}
\end{figure*}

% Since the cross-layer data collected on the 5G NSA testbed is inherently temporally correlated, it is essential to utilize this temporal correlation between the successive data samples.
% Sequential models that can utilize this temporal correlation, e.g., the LSTM and the ESN, are also implemented. 
Next, the jamming detection performance is evaluated with the sequential models described in Sec.~\ref{sec:machinelearningmodel}. 
We use $T=2$ consecutive data samples as described in Eq.~\eqref{eq:temporal_data} to construct the ``sequence'' of feature data that is input to the model, with the first data sample (feature vector) in each sequence belonging to the 5G NR cell and the other sample belonging to the LTE cell. 
% For sequential learning, we consider machine learning models belonging to the recurrent neural network (RNN) family, namely the Long Short-Term Memory (LSTM) and the Echo State Network (ESN), as outlined previously in Sec.~\ref{sec:machinelearningmodel}. 
The detailed training setup for each model is provided in Table~\ref{tab:training_setup_detection}.
For each model, the Adam optimizer was used with a learning rate of $0.001$ and a batch size of $32$.
The ROC curve performance for the LSTM jamming detector in different jamming scenarios is shown in Fig.~\ref{roc_lstm}. The detailed performance metrics of the LSTM-based detector along with those of the GRU-based, CNN-LSTM based and Transformer-based detectors are also given in Table~\ref{tab:jamming_detection_performance}.

\begin{table}[!h]
\centering
\caption{Training Setup for the Time Series-based models for Jamming Detection}
\label{tab:training_setup_detection}
\begin{tabular}{@{}ccc@{}}
\toprule
\textbf{Model Type} & \textbf{Trainable Parameters} & \multicolumn{1}{l}{\textbf{Training Epochs}} \\ \midrule
LSTM        & 4,326   & 30 \\
GRU         & 3,326   & 30 \\
CNN-LSTM    & 24,706  & 25 \\
Transformer & 292,553 & 50 \\ \bottomrule
\end{tabular}
\end{table}

% \begin{table}[!h]
% \centering
% \caption{Model Parameter and Hyperparameter Settings for the Transformer and Bidirectional ESN models}
% \label{tab:xfmr_esn_hyperparameters}
% \begin{tabular}{cc}
% \hline
% \textbf{Model Type} &
%   \textbf{Model Parameters} \\ \hline
% Transformer &
%   \begin{tabular}[c]{@{}l@{}}\textbf{Multi-Head Attention in Transformer Encoder}:\\ Number of heads = 4\\ Head size = 256\\  \textbf{Feedforward network in Transformer Encoder:}\\ 1D convolutional layer with filters = 4\end{tabular} \\ \hline
% Bidirectional ESN &
%   \begin{tabular}[c]{@{}l@{}}Reservoir units = 125\\ Spectral radius = 0.45\\ Leakage rate ($\alpha$) = 0.1\\ Connectivity = 0.4\\ Trainable parameters (bidirectional ESN): \\ Detection: 250, Classification: 1000\end{tabular} \\ \hline
% \end{tabular}
% \end{table}

\begin{table}[!h]
\caption{Model Parameter and Hyperparameter Settings for the Transformer and Bidirectional ESN models}
\label{tab:xfmr_esn_hyperparameters}
\begin{tabular}{cl}
\hline
\textbf{Model Type} &
  \multicolumn{1}{c}{\textbf{Model Parameters}} \\ \hline
Transformer &
  \begin{tabular}[c]{@{}l@{}}\textbf{Multi-Head Attention (MHA) in Transformer encoder:}\\ Number of heads = $4$\\ Head size = $256$\\ \textbf{Feedforward network in Transformer encoder:}\\ 1D convolutional layer with filters = $4$\end{tabular} \\ \hline
\begin{tabular}[c]{@{}c@{}}Bidirectional \\ ESN\end{tabular} &
  \begin{tabular}[c]{@{}l@{}}Reservoir units ($M$) = $125$\\ Spectral radius ($\rho$) = $0.45$\\ Leakage ($\alpha$) = $0.9$\\ Connectivity ($\kappa$) = $0.4$\\
  Regularization constant ($\lambda$) = $0.01$ \\\textbf{Number of trainable parameters (Bidirectional ESN):}\\ Detection: $250$, Classification: $1000$\end{tabular} \\ \hline
\end{tabular}
\end{table}

% Our LSTM implementation uses a single hidden layer consisting of $50$ LSTM units and an optional Dropout layer~\cite{Dropout2014}, appended with a final output layer with $2$ nodes for binary jamming detection. Threshold adjusting using Youden's J statistic is also employed to significantly improve detection accuracy as well as the precision and recall performance. 

\begin{figure*}[h]
\centering
\includegraphics[width=0.95\textwidth]{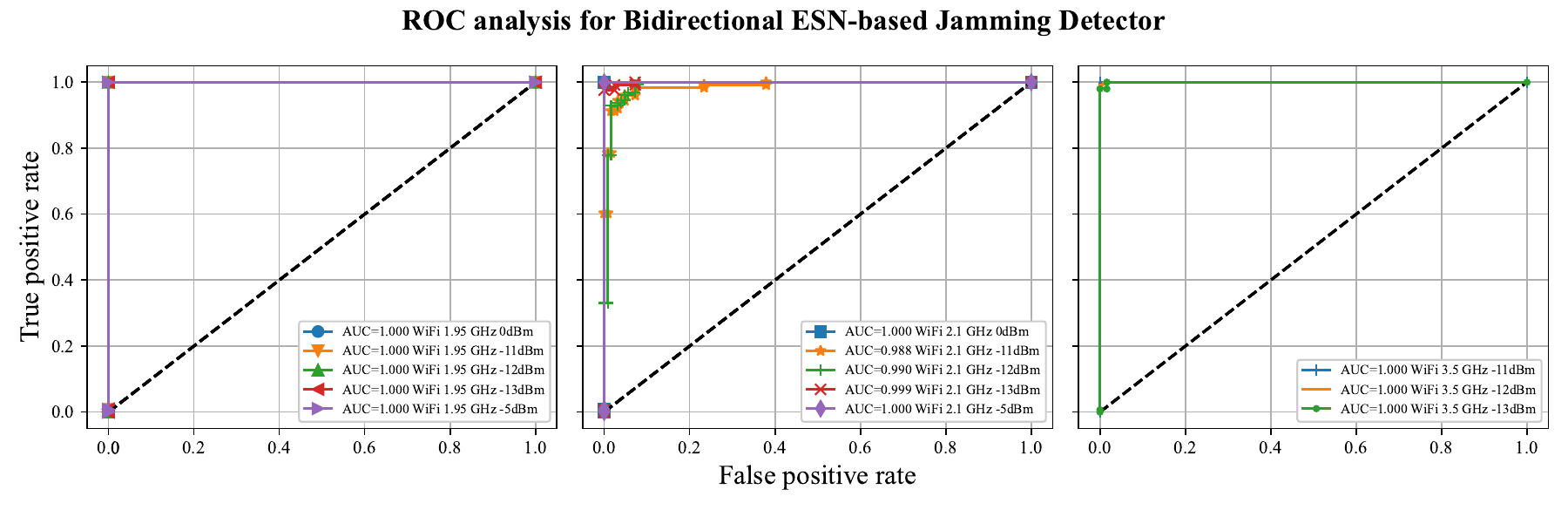}
\caption{Receiver Operating Characteristic (ROC) curves for each jamming scenario with the sequential bidirectional RC (ESN)-based detector.}
\label{roc_esn}
\end{figure*}

Finally, we also implement the bidirectional RC (ESN)-based detector with its detailed setup outlined in Table~\ref{tab:xfmr_esn_hyperparameters}.
% with its reservoir size, i.e., number of neurons in the reservoir set to $50$, in order to keep its overall same model size (albeit with significantly lower trainable parameters) comparable with the LSTM detector. 
The corresponding ROC curves for the ESN detector is shown in Fig.~\ref{roc_esn} with its other performance metrics outlined in Table~\ref{tab:jamming_detection_performance}. 
We can see that the performance of the bidirectional ESN detector is comparable to that of all of the standard recurrent detectors, e.g., the accuracy of the ESN detector is only $6.8\%$ lower compared to the best performing GRU detector and has a $27\%$ higher False Alarm Rate compared to the GRU.
However, as we shall see in Sec.~\ref{sec:bayesian_results}, Bayesian causal inference built on domain knowledge can significantly improve the accuracy and lower the FAR.
% \textcolor{blue}{As a final observation, we can see across both categories of detectors that detecting jamming signals is relatively more difficult in the LTE DL band ($2.14$ GHz) as compared to the LTE UL band ($1.95$ GHz) in a legitimate 5G NR link.}
% the LSTM detector under interference scenarios in all three bands but with significantly lower training and test complexity.

% This demonstrates that a lightweight model such as the ESN has great applicability in IoT applications under a near real-time performance constraint, when there is substantial benefit to be gained from exploiting temporal correlation in the sensed or generated data.

% Please add the following required packages to your document preamble:
% \usepackage{booktabs}
% \usepackage{multirow}
\begin{table*}[h]
\centering
\caption{Jamming Detection Performance Evaluation (Averaged across 13 Jamming Scenarios)}
\label{tab:jamming_detection_performance}
\begin{tabular}{@{}clcccccc@{}}
\toprule
\textbf{Model Type} &
  \multicolumn{1}{c}{\textbf{Model Name}} &
  \textbf{Avg AUC} &
  \textbf{Accuracy} &
  \textbf{Precision} &
  \textbf{Recall} &
  \multicolumn{1}{l}{\textbf{F-score}} &
  \multicolumn{1}{l}{\textbf{False Alarm Rate}} \\ \midrule
\multirow{6}{*}{Instantaneous} & Random   Forest        & 0.9887 & 0.9537 & 0.9558 & 0.9539 & 0.9549 & 0.0466 \\
                               & AdaBoost               & 0.9840 & 0.9459 & 0.9457 & 0.9476 & 0.9467 & 0.0559 \\
                               & Logistic Regression    & 0.9549 & 0.9019 & 0.8992 & 0.9064 & 0.9028 & 0.1026 \\
                               & $K$-Nearest Neighbors            & 0.7396 & 0.6769 & 0.6778 & 0.6791 & 0.6785 & 0.3254 \\
                               & Gaussian Naive Bayes            & 0.6750 & 0.6253 & 0.6321 & 0.6346 & 0.6333 & 0.3845 \\
                               & Bagging classifier     & 0.9839 & 0.9492 & 0.9535 & 0.9440 & 0.9487 & 0.0457 \\ \midrule
\multirow{5}{*}{\begin{tabular}[c]{@{}c@{}}Time \\ series-based\end{tabular}} &
  LSTM &
  0.9979 &
  0.9920 &
  0.9948 &
  0.9896 &
  0.9922 &
  0.0055 \\
                               & GRU                    & 0.9996 & 0.9945 & 0.9975 & 0.9915 & 0.9945 & 0.0025 \\
                               & CNN-LSTM               & 0.9994 & 0.9938 & 0.9979 & 0.9900 & 0.9939 & 0.0022 \\
                               & Transformer            & 0.9692 & 0.9543 & 0.9624 & 0.9509 & 0.9552 & 0.0416 \\
                               & Bidirectional RC (ESN) & 0.9982 & 0.9878 & 0.9913 & 0.9848 & 0.9879 & 0.0093 \\ \bottomrule
\end{tabular}
\end{table*}

% \begin{figure}[!h]
% \centering
% \includegraphics[width=0.9\textwidth]{FIGURES_FINAL/ESN/roc_esn_040423.png}
% \caption{The Overview of AUC for Each Jamming Type with ESN-based detector.}
% \label{roc_esn}
% \end{figure}

\begin{figure*}[!h]
    \centering
    \subfloat[t-SNE: Instantaneous Detector.]{\includegraphics[width=0.275\linewidth, height=0.275\linewidth]{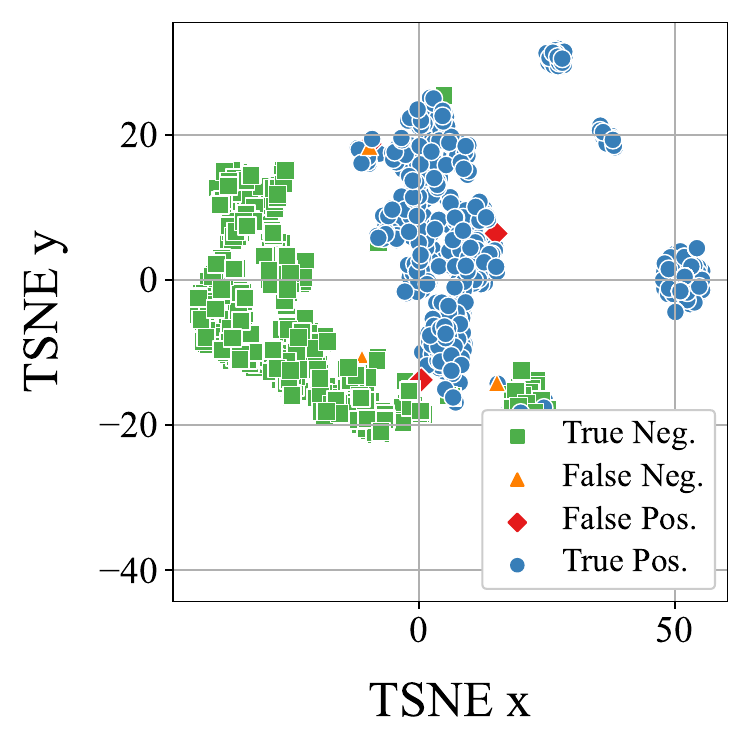}\label{fig:inst_tsne}}%
    \subfloat[t-SNE: LSTM-based Detector.]{\includegraphics[width=0.275\linewidth, height=0.275\linewidth]{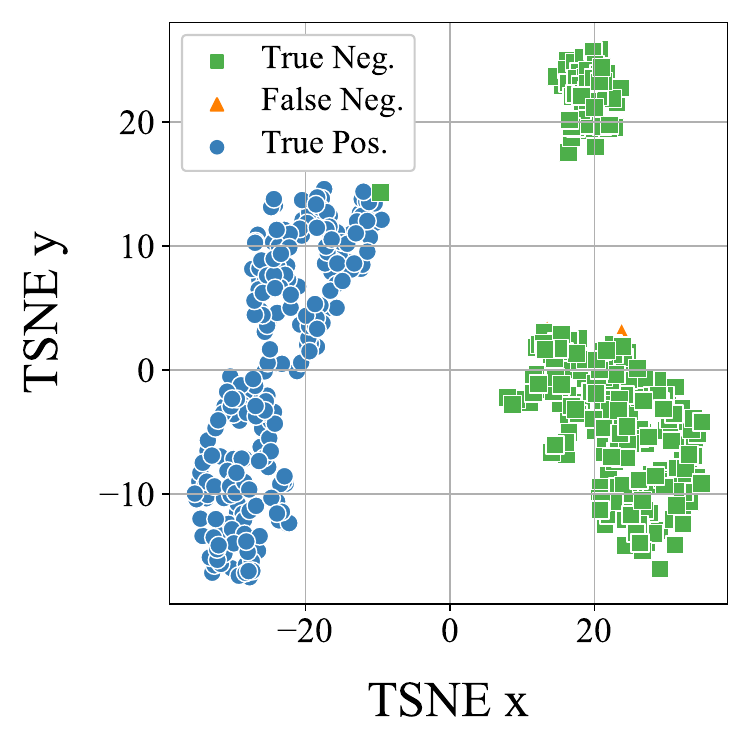}\label{fig:lstm_tsne}}%
    \subfloat[t-SNE: Bidirectional ESN-based Detector.]{\includegraphics[width=0.275\linewidth, height=0.275\linewidth]{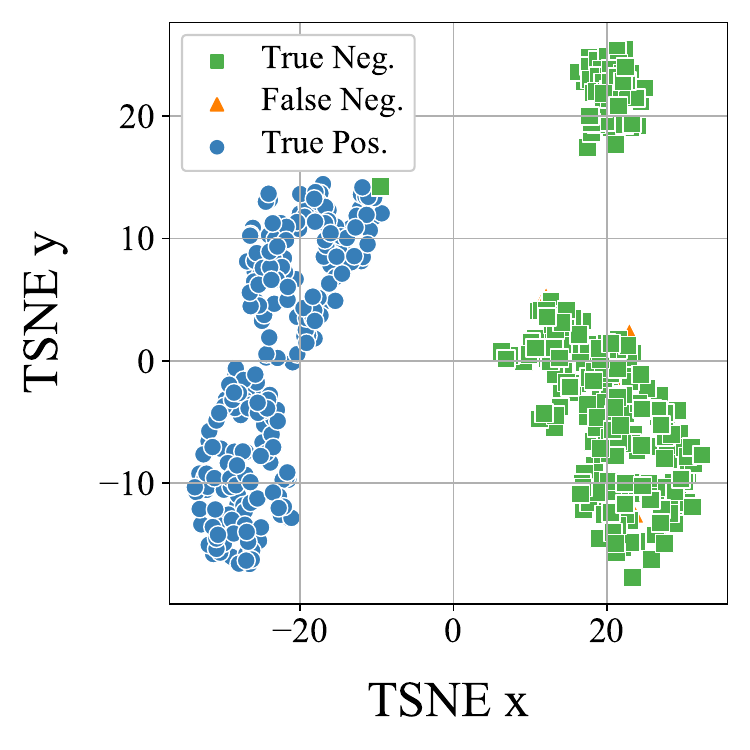}\label{fig:esn_tsne}}%
    \caption{t-SNE plots comparing the Instantaneous (Random Forest) detector versus two Sequential time series-based models.}
\end{figure*}

In order to obtain an alternate viewpoint of the effectiveness of the LSTM and ESN models processing sequential inputs in discriminating between the feature space of TPs, TNs, FPs and FNs, we also employ the t-distributed Stochastic Neighbor Embedding (t-SNE) visualization method. 
The t-SNE~\cite{vandermaaten08a} is a method of transforming high-dimensional data into a two or three-dimensional map for easier visualization. In our setup, the original high-dimensional data is condensed into a flat two-dimensional map.
The input parameters used to create the two-dimensional t-SNE plots include the CQI, UL MCS, DL MCS and the PUSCH SNR, while the outputs are the four Confusion Matrix (CM) elements.
Specifically, the four CM elements are plotted in this reduced two-dimensional space following dimensionality reduction from the original four-dimensional space.
In the t-SNE plots of Fig.~\ref{fig:inst_tsne}, we can see that for the instantaneous random forest model, the clusters of TPs and TNs are mostly distinguishable with only a few outliers (mis-clustering). 
Similarly, in the t-SNE plots for the LSTM and ESN models in Fig.~\ref{fig:lstm_tsne} and Fig.~\ref{fig:esn_tsne} respectively, note again that the clusters of TPs and TNs are clearly separable. However, the TP and TN points themselves do not exhibit multi-level clustering to the same extent as the instantaneous model, indicating easier identification of the incorrect predictions (FPs and FNs) in this reduced two-dimensional space.
% separation is more pronounced in the case of the time series-based LSTM and ESN models with only a single cluster for either TP or TN. 
This demonstrates the value of utilizing the temporal correlation inherent in the data, which the LSTM and ESN models can exploit whereas the discriminative instantaneous model does not, which is evident from the degree of separation of the four CM element clusters in their respective t-SNE plots. 
% The inherent benefit of utilizing this correlation becomes more pronounced as the training dataset increases in size.

% \begin{figure}[!h]
%     \centering
%     \includegraphics[width=0.3\textwidth]{Maj_Rev_Figs/INSTA_TSNE.eps}
%     \caption{t-SNE: Instantaneous model}
%     \label{fig:inst_tsne}
% \end{figure}

% \begin{figure}[!h]
%     \centering
%     \includegraphics[width=0.3\textwidth]{Maj_Rev_Figs/LSTM_TSNE.eps}
%     \caption{t-SNE: LSTM model}
%     \label{fig:lstm_tsne}
% \end{figure}

% \begin{figure}[!h]
%     \centering
%     \includegraphics[width=0.3\textwidth]{Maj_Rev_Figs/ESN_TSNE.eps}
%     \caption{t-SNE: ESN Model}
%     \label{fig:esn_tsne}
% \end{figure}

\subsection{Multi-Scenario Jamming Classification Performance}
\label{sec:multi_scenario_classification_performance_evaluation}
\begin{figure}[h]
    \centering    \includegraphics[width=0.9\linewidth]{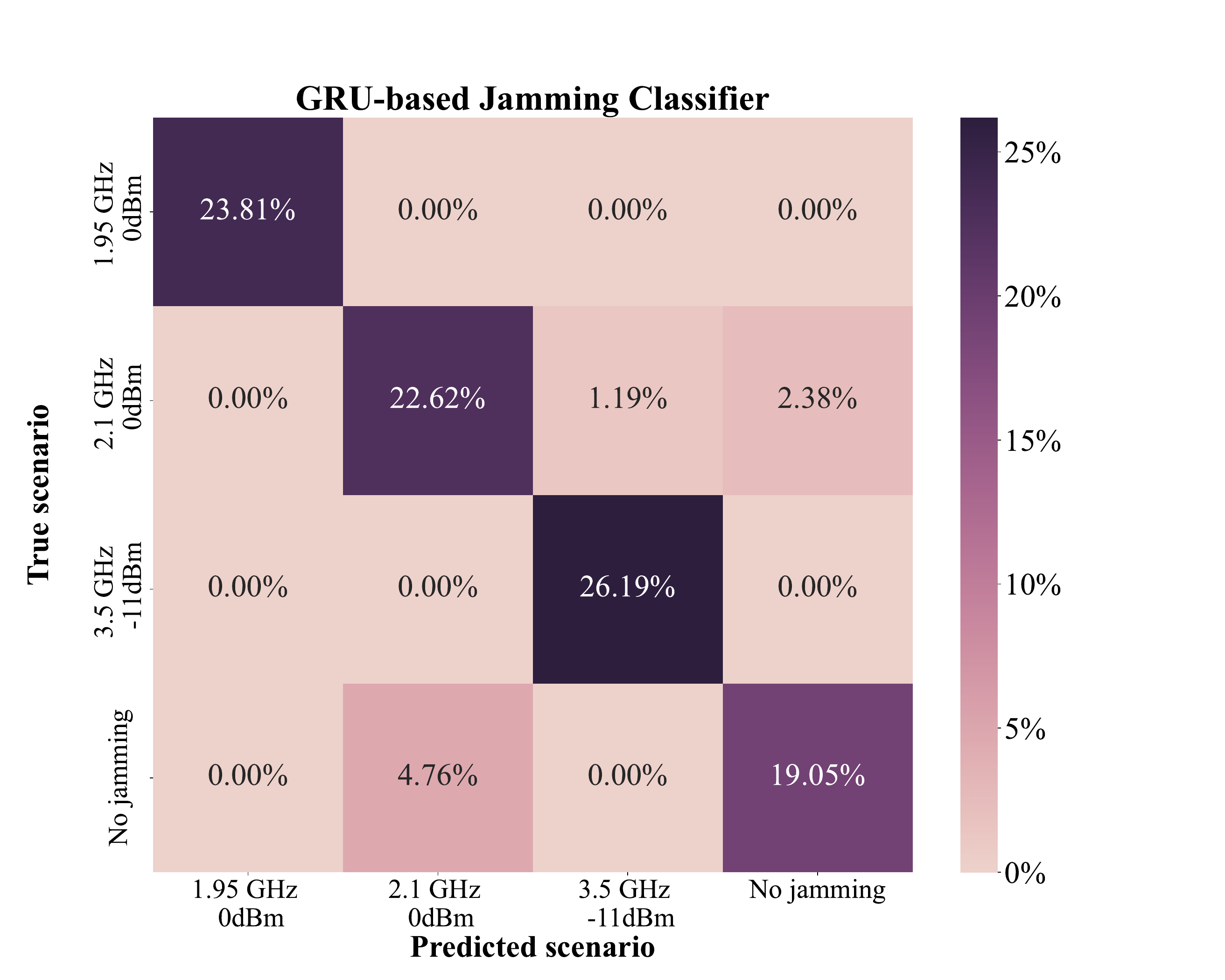}
    \caption{Confusion Matrix for the GRU-based classifier.}
    \label{fig:gru_confusion_matrix_classification}
\end{figure}

\begin{figure}[h]
    \centering
    \includegraphics[width=0.9\linewidth]{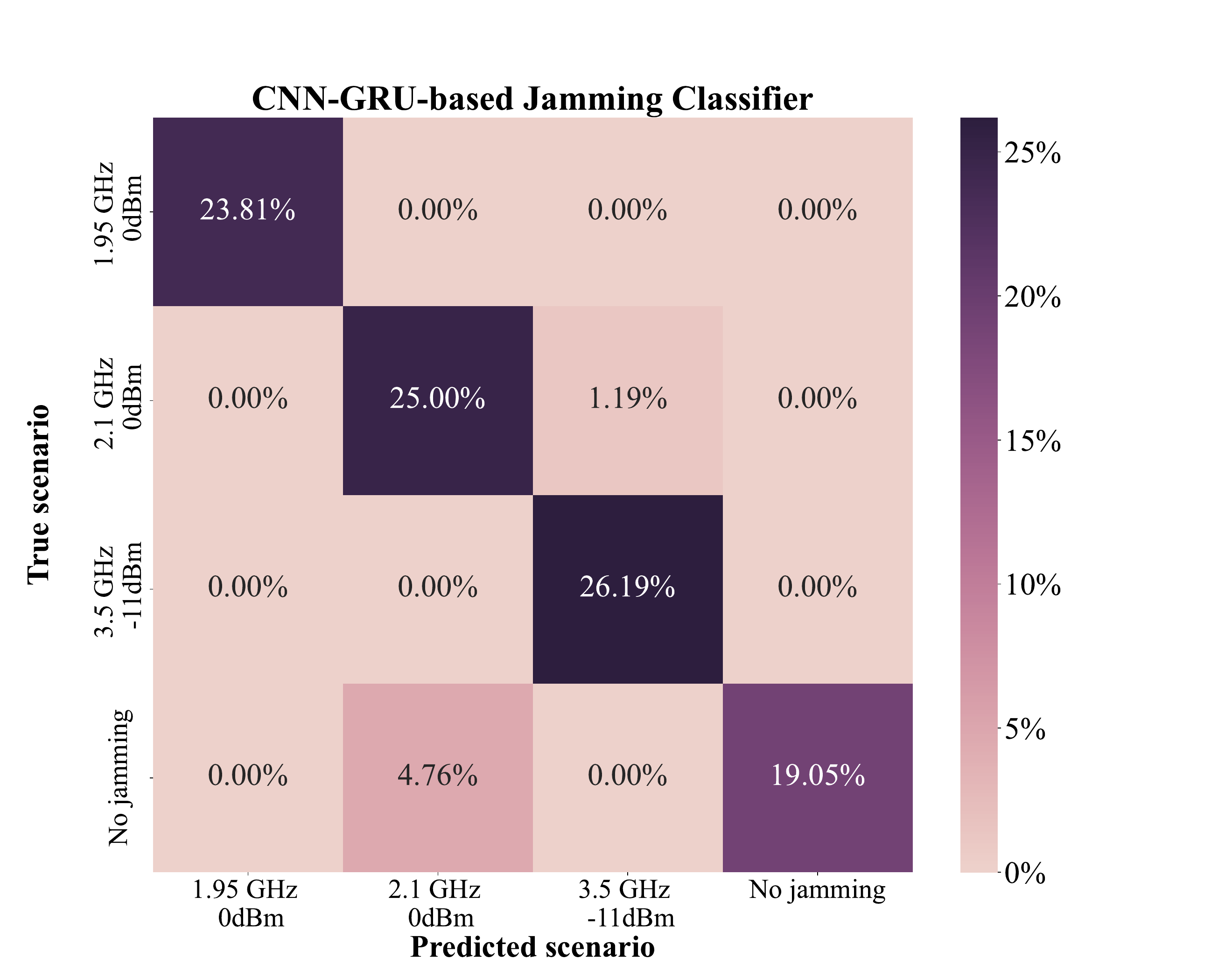}
    \caption{Confusion Matrix for the CNN-GRU based classifier.}    \label{fig:cnn_gru_confusion_matrix_classification}
\end{figure}

\begin{figure}[h]
    \centering
    \includegraphics[width=0.9\linewidth]{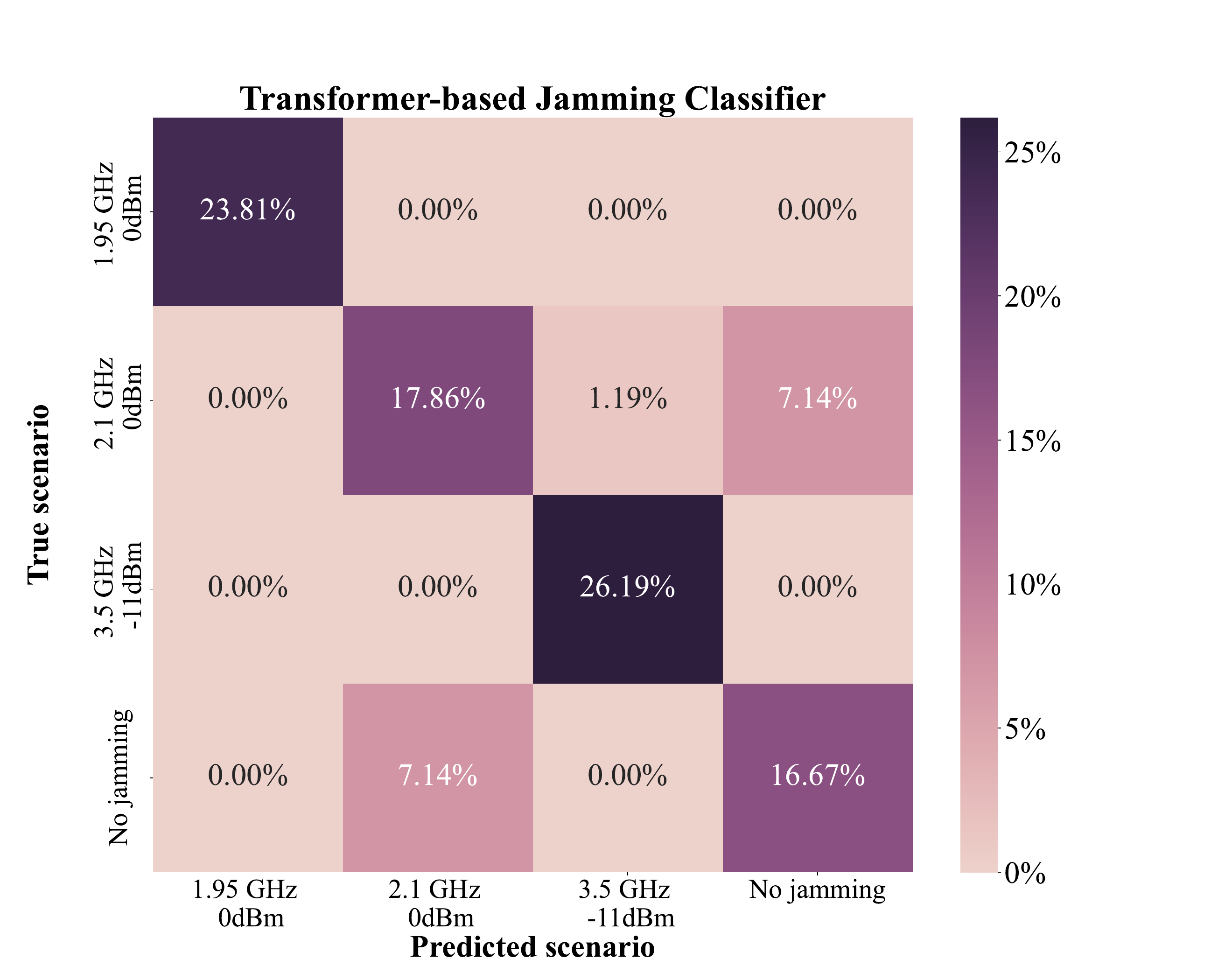}
    \caption{Confusion Matrix for the Transformer-based classifier.}    \label{fig:xfmr_confusion_matrix_classification}
\end{figure}

\begin{figure}[h]
    \centering
    \includegraphics[width=0.9\linewidth]{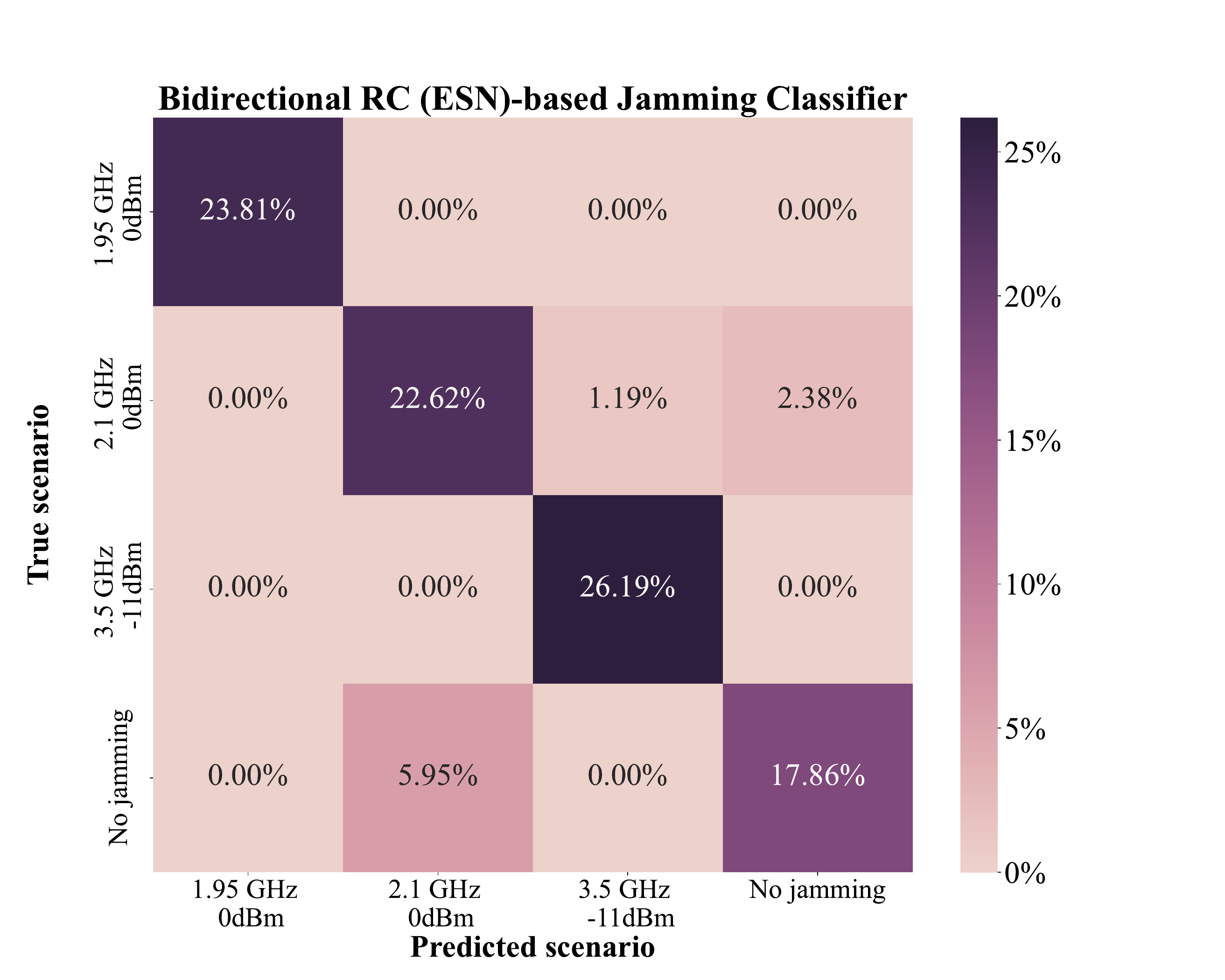}
    \caption{Confusion Matrix for the Bidirectional RC (ESN)-based classifier.}
    \label{fig:esn_confusion_matrix_classification}
\end{figure}

% \textcolor{blue}{Add Confusion matrices}
Following the description in Sec.~\ref{sec:jamming_classification_description} for models employed to classify between different jamming scenarios characterized by distinct center frequencies, we provide a complete performance evaluation for the same in this section. 
The training setup for the sequential models considered under the multi-scenario classification task is summarized in Table~\ref{tab:training_setup_classification}.
\begin{table}[!h]
\centering
\caption{Training Setup for the Time Series-based models
for Multi-Scenario Classification}
\label{tab:training_setup_classification}
\begin{tabular}{@{}ccc@{}}
\toprule
\textbf{Model Type} & \textbf{Trainable Parameters} & \multicolumn{1}{l}{\textbf{Training Epochs}} \\ \midrule
GRU         & 46,204  & 25 \\
CNN-GRU     & 75,984  & 25 \\
Transformer & 295,500 & 45 \\ \bottomrule
\end{tabular}
\end{table}
The confusion matrices for 
% both the instantaneous random forest as well as 
the sequential models considered are depicted in
% Fig.~\ref{fig:inst_confusion_matrix_classification} and 
Fig.~\ref{fig:gru_confusion_matrix_classification}, Fig.~\ref{fig:cnn_gru_confusion_matrix_classification}, Fig.~\ref{fig:xfmr_confusion_matrix_classification} and Fig.~\ref{fig:esn_confusion_matrix_classification} respectively.
% \begin{figure}[h]
%     \centering
%     \includegraphics[width=0.975\linewidth]{Maj_Rev_Figs/Inst_Multi-Scenario_CM_V2.eps}
%     \caption{Confusion Matrix for the Instantaneous Random Forest classifier.}    \label{fig:inst_confusion_matrix_classification}
% \end{figure}
We can observe that the random forest classifier which only works on a single time step of data performs considerably poorly with a high percentage of misclassification errors compared to the sequential models, signifying the importance of exploiting any available temporal correlation embedded in the input sequence.
The confusion matrix in Fig.~\ref{fig:esn_confusion_matrix_classification} for the bidirectional ESN classifier shows superior performance, comparable to the GRU and CNN-GRU based classifiers.
Similar to the detection problem, the transformer's confusion matrix in Fig.~\ref{fig:xfmr_confusion_matrix_classification} depicts overfitting on the limited training data leading to slightly worse misclassification results.
The complete performance characterization for multi-scenario classification is summarized in Table~\ref{tab:multi_scenario_classification_performance_evaluation}.

\begin{table*}[!h]
\centering
\caption{Multi-Scenario Classification Performance Evaluation (Averaged across 4 Jamming Scenarios)}
\label{tab:multi_scenario_classification_performance_evaluation}
\begin{tabular}{@{}llcccc@{}}
\toprule
\multicolumn{1}{c}{\textbf{Model Type}} &
  \multicolumn{1}{c}{\textbf{Model Name}} &
  \textbf{Accuracy} &
  \textbf{Precicion} &
  \textbf{Recall} &
  \multicolumn{1}{l}{\textbf{F-score}} \\ \midrule
% \multicolumn{1}{c}{Instantaneous}  & Random   Forest        & 0.5680 & 0.6190 & 0.5764 & 0.5447 \\ \midrule
\multirow{4}{*}{Time series-based} & GRU                    & 0.9167 & 0.9179 & 0.9159 & 0.9161 \\
                                   & CNN-GRU                & 0.9405 & 0.9491 & 0.9386 & 0.9401 \\
                                   & Transformer            & 0.8452 & 0.8427 & 0.8455 & 0.8439 \\
                                   & Bidirectional RC (ESN) & 0.9048 & 0.9076 & 0.9034 & 0.9037 \\ \bottomrule
\end{tabular}
\end{table*}

\subsection{Training Details and Run-time Evaluation}
\begin{figure}[h]
    \centering
    \includegraphics[width=\linewidth]{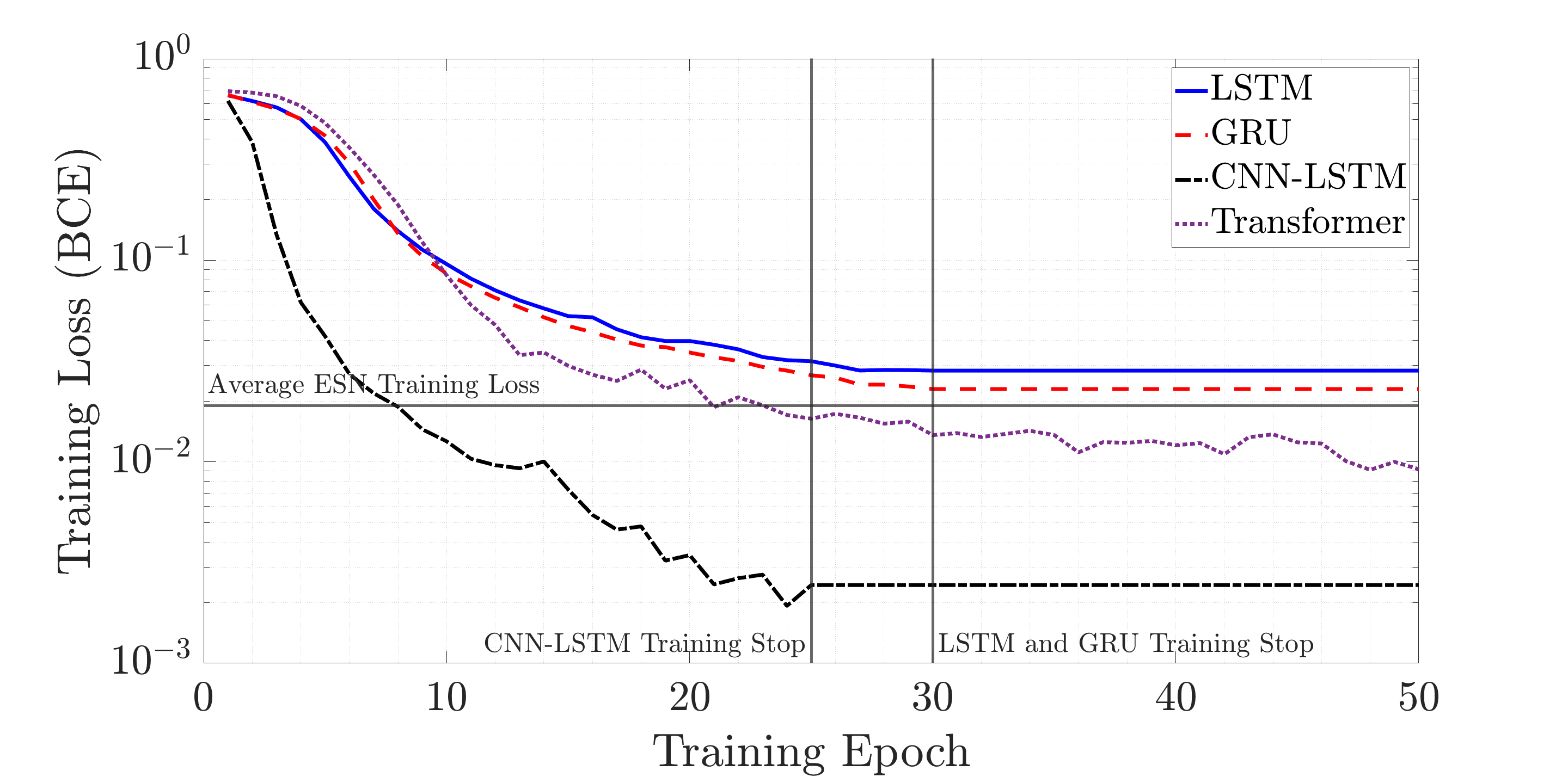}
    \caption{Convergence of training loss (BCE) for the jamming detection task. Average ESN training loss shown for comparison.}
    \label{fig:training_loss_convergence_detection}
\end{figure}

\begin{figure}[h]
    \centering
    \includegraphics[width=\linewidth]{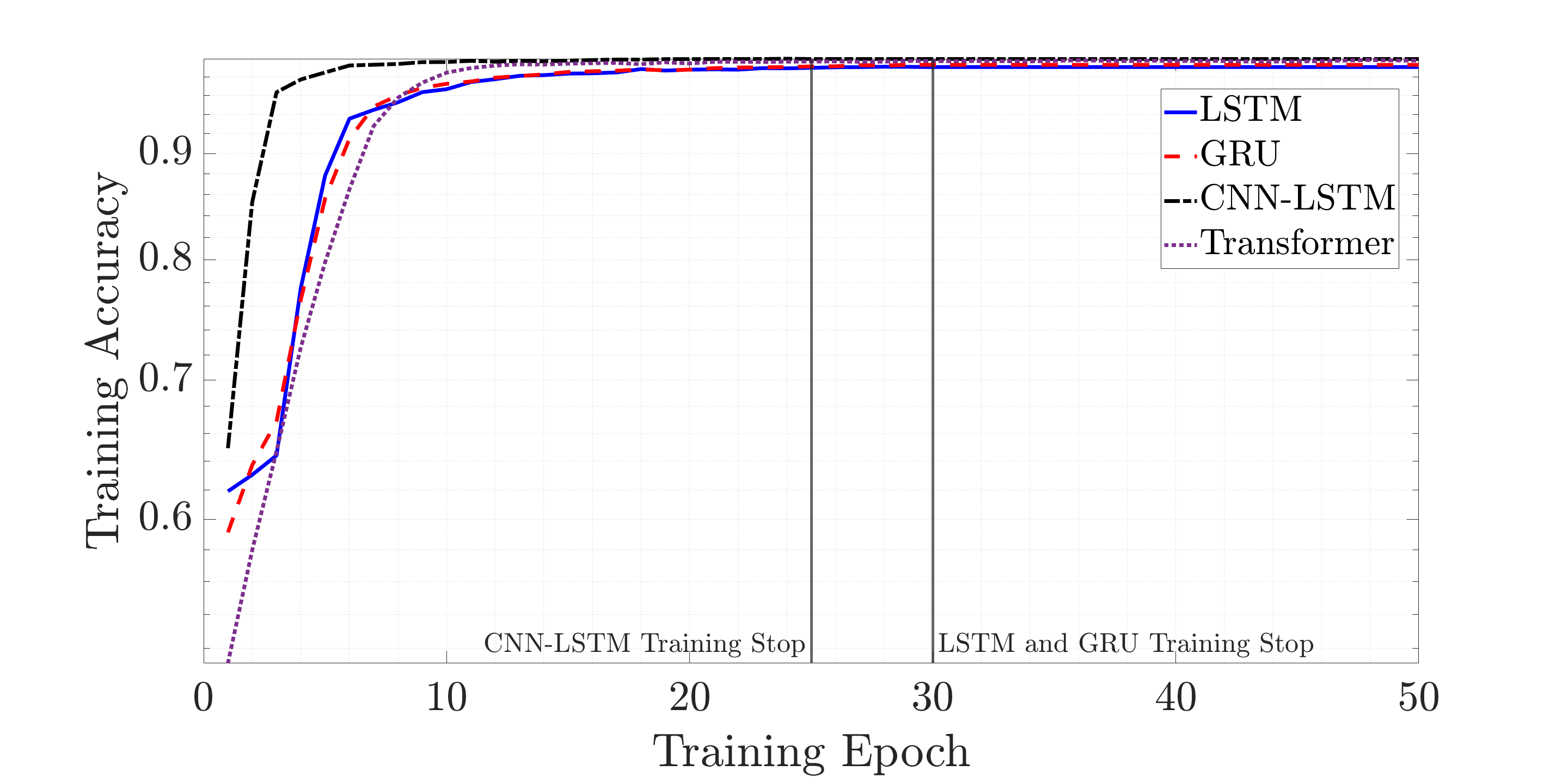}
    \caption{Convergence of training accuracy for the jamming detection task.}
    \label{fig:training_accuracy_convergence_detection}
\end{figure}

In this section, we provide details on the convergence observed during training of the sequential neural network models trained using backpropagation through time (BPTT).
For the binary detection task, the loss metric used is the binary cross entropy (BCE). The BCE loss as a function of the training epoch is shown in Fig.~\ref{fig:training_loss_convergence_detection} while the training accuracy versus training epoch is shown in Fig.~\ref{fig:training_accuracy_convergence_detection}.
The number of training epochs for each model is based on the trend observed in their corresponding validation losses, when a validation split of $20\%$ was used.
% The overall setup for the training of the time series-based models under the binary jamming detection is provided in Table~\ref{tab:training_setup_detection}. For each model, the Adam optimizer was used with a learning rate of $0.001$ and a batch size of $32$.
Next, the detailed train and test BCE loss metrics for the jamming detection task as well as the CPU train and test run times for all models are given in Table~\ref{tab:loss_metrics_run_times_jamming_detection}\footnote{The hardware used was an Intel Core i7-1260U processor running at 3.4 GHz with a memory of 32 GB.}.
Note that the bidirectional ESN detector has an average CPU train time of approximately $53$ ms and an average CPU inference time of approximately $2$ ms.
This makes it feasible to implement an RC-based jamming detector with near real-time training and inference capabilities, similar to our previous work in~\cite{Li2020wifi} and~\cite{Liang2022} that implements an ESN on a software defined radio (SDR) transceiver for real-time MIMO-OFDM symbol detection.
\begin{table*}[htbp]
\centering
\caption{Training and Test Loss metrics and CPU Run Times for Jamming Detection (Averaged across 13 Jamming Scenarios)}
\label{tab:loss_metrics_run_times_jamming_detection}
\begin{tabular}{@{}cccccc@{}}
\toprule
\textbf{Model Type} &
  \multicolumn{1}{c}{\textbf{Model Name}} &
  \textbf{Avg Train Loss (BCE)} &
  \textbf{Avg Test Loss (BCE)} &
  \textbf{Avg Train Time (sec)} &
  \multicolumn{1}{l}{\textbf{Avg Test Time (sec)}} \\ \midrule
\multirow{6}{*}{Instantaneous} & Random   Forest        & 0.0355 & 0.1209 & 0.4442  & 0.0132 \\
                               & AdaBoost               & 0.5257 & 0.5261 & 0.2140  & 0.0074 \\
                               & Logistic Regression    & 0.2175 & 0.2383 & 0.9039  & 0.0003 \\
                               & K-Neighbors            & 0.4147 & 1.9115 & 0.0016  & 0.1809 \\
                               & Gaussian NB            & 0.6174 & 0.6194 & 0.0010  & 0.0002 \\
                               & Bagging classifier     & 0.0319 & 0.2719 & 0.2344  & 0.0030 \\ \midrule
\multirow{5}{*}{\begin{tabular}[c]{@{}c@{}}Time \\ series-based\end{tabular}} &
  LSTM &
  0.0283 &
  0.4431 &
  6.8611 &
  0.1219 \\
                               & GRU                    & 0.0248 & 0.4694 & 6.2089  & 0.1081 \\
                               & CNN-LSTM               & 0.0215 & 0.4626 & 7.5704  & 0.1227 \\
                               & Transformer            & 0.0142 & 1.0805 & 56.9653 & 0.3359 \\
                               & Bidirectional RC (ESN) & 0.0128 & 0.4127 & 0.0527  & 0.0017 \\ \bottomrule
\end{tabular}
\end{table*}
Similarly, the train and test categorical cross entropy (CCE) losses for the multi-scenario classification task are shown in given in Table~\ref{tab:loss_metrics_run_times_multi_scenario_classification}.
Similar to the detection task, the ESN-based classifier exhibits a CPU training run time of approximately $30$ ms and a CPU inference run time of approximately $1.5$ ms, making possible its implementation on IoT devices that may have much lower processor clock frequencies and still satisfy the $180$ ms (or similar) inference time criterion.
Note that the relatively higher average test CCE loss for the bidirectional ESN classifier in Table~\ref{tab:loss_metrics_run_times_multi_scenario_classification} is likely due to the model size and capacity not being large enough for a $4$-class classification task given the size of the dataset being used for training.
However, the hard decision metrics such as accuracy are not adversely affected despite the higher test CCE loss.

\begin{table*}[!h]
\centering
\caption{Training and Test Loss Metrics and CPU Run Times for Multi-Scenario Classification (Averaged across 4 Jamming Scenarios)}
\label{tab:loss_metrics_run_times_multi_scenario_classification}
\begin{tabular}{@{}llcccc@{}}
\toprule
\multicolumn{1}{c}{\textbf{Model Type}} &
  \multicolumn{1}{c}{\textbf{Model Name}} &
  \multicolumn{1}{l}{\textbf{Avg train loss (CCE)}} &
  \multicolumn{1}{l}{\textbf{Avg test loss (CCE)}} &
  \multicolumn{1}{l}{\textbf{Avg train time (sec)}} &
  \multicolumn{1}{l}{\textbf{Avg Test time (sec)}} \\ \midrule
% \multicolumn{1}{c}{Instantaneous}  & Random   Forest        & 0.0424 & 0.9634 & 1.4045   & 0.0207 \\ \midrule
\multirow{4}{*}{Time series-based} & GRU                    & 0.0056 & 0.3020 & 4.0416   & 0.0490 \\
                                   & CNN-GRU                & 0.0241 & 0.3564 & 5.4705   & 0.0651 \\
                                   & Transformer            & 0.0012 & 0.6709 & 102.4488 & 0.2413 \\
                                   & Bidirectional RC (ESN) & 0.0286 & 1.4123 & 0.0295   & 0.0014 \\ \bottomrule
\end{tabular}
\end{table*}

\subsection{Bayesian Causal Analysis for Jamming Detection}
\label{sec:bayesian_results}
As mentioned in Sec.~\ref{sec:causal}, a BNM is effective when the number of data samples available is insufficient to train black-box ML models that have a high model capacity, to the effect that overfitting may occur. In this section, we present results for the BNM causal analysis conducted on specific interference scenarios where the performance suffers due to fewer data samples being available than required and thereby manifests as relatively poor detection AUC in both instantaneous as well as sequential time series-based detection modules.
Accordingly, the training-test data split is now reduced to $50-50 \%$, so that all deep learning models experience performance degradation.
For the time series-based detection models, we consider the scenario of jamming present in the LTE DL channel ($2.14$ GHz) of band $B1$ at a power level of $-13$ dBm. 

\begin{figure}[t]
\centering
\includegraphics[width=0.4875\textwidth]{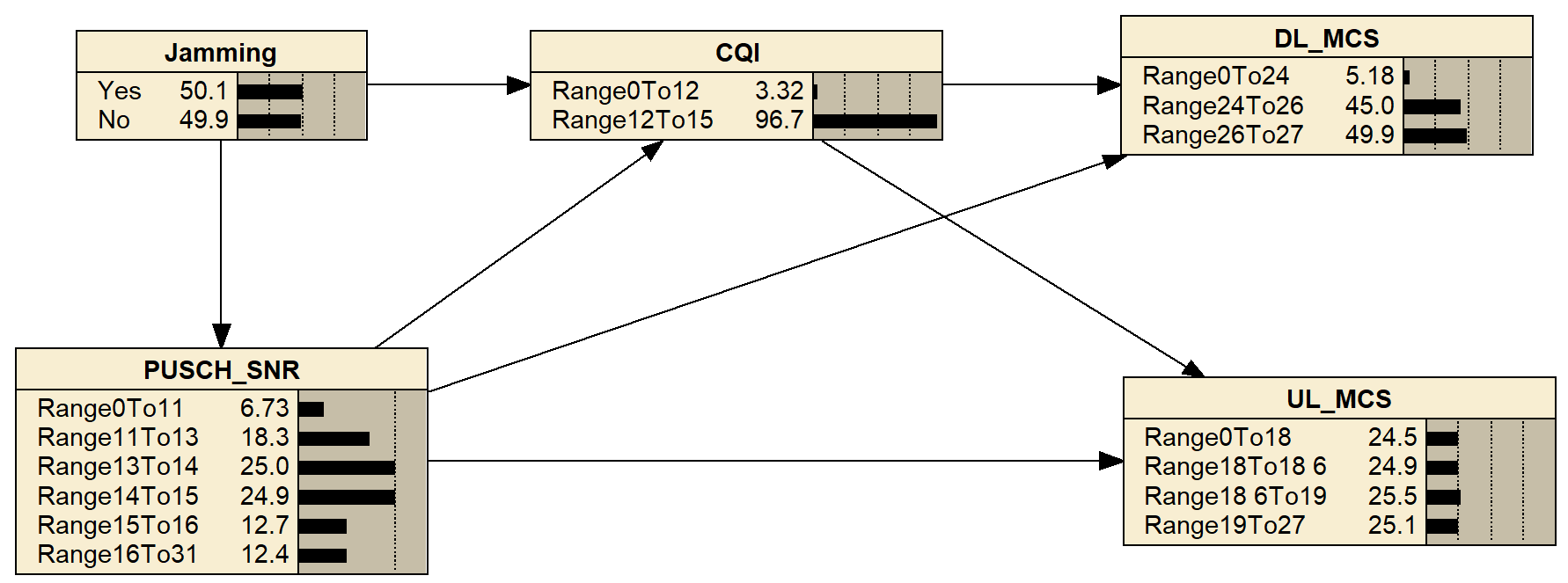}
\caption{Bayesian Network Model (BNM) for jamming in the LTE DL band (2.1 GHz) at -13 dBm.}
\label{fig:bnm_2p1GHz_m13dBm}
\end{figure}

\begin{figure}[t]
\centering
\includegraphics[width=0.4875\textwidth]{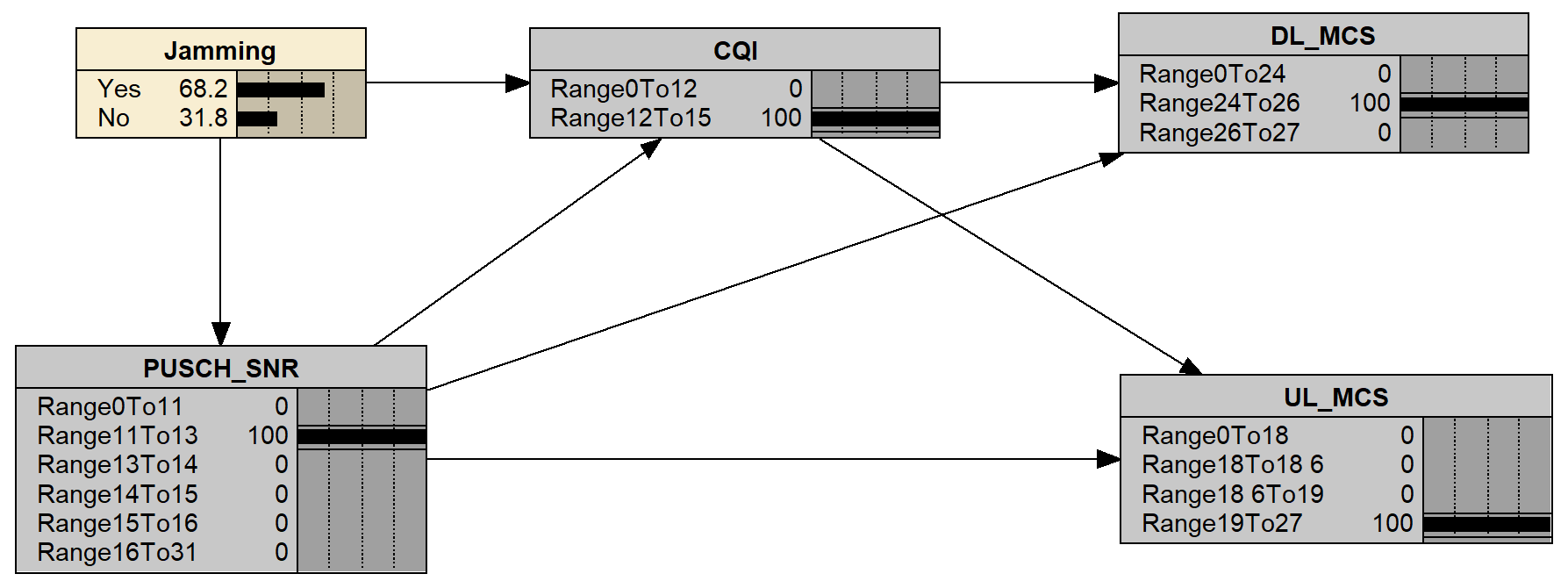}
\caption{BNM posterior probability for jamming in the LTE DL band (2.1 GHz) at -13 dBm for False Negative ESN predictions. A grayed node represents knowledge of the observed value of that (random) variable with the value at each node specified by the corresponding range selected.}
\label{fig:bnm_2p1GHz_m5dBm_lstm_fn}
\end{figure}

\begin{figure}[t]
\centering
\includegraphics[width=0.4875\textwidth]{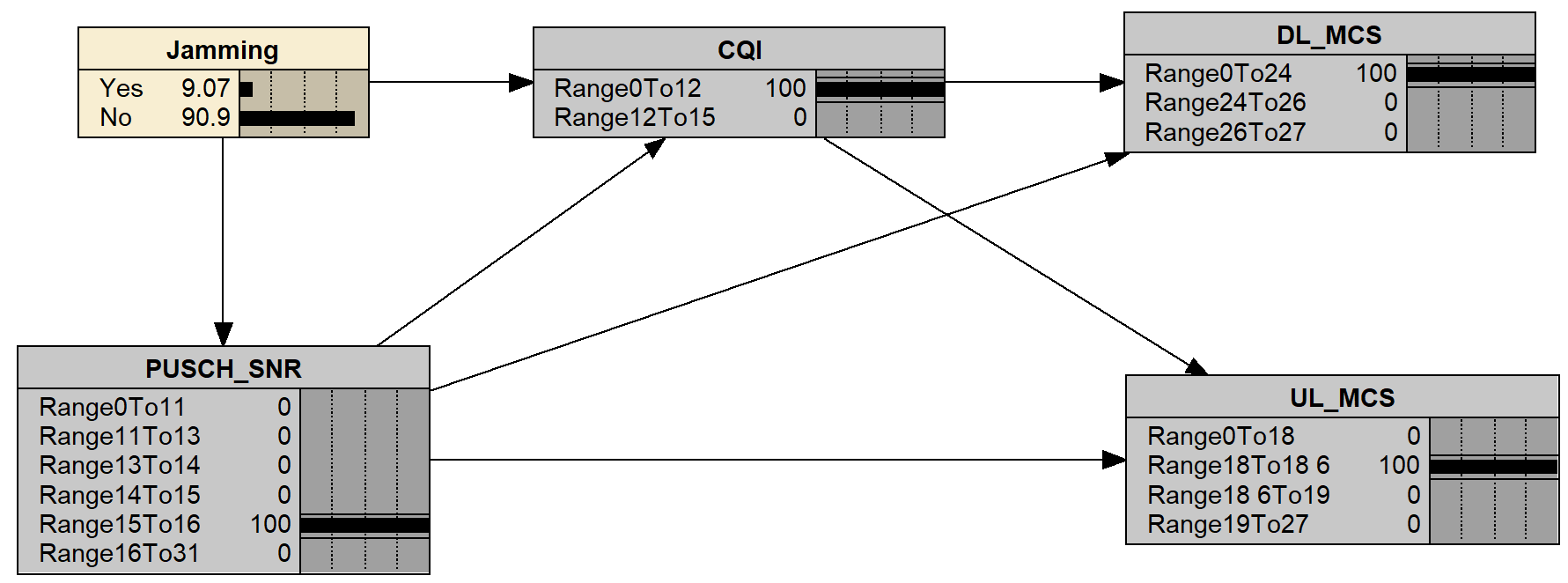}
\caption{BNM posterior probability for jamming in the LTE DL band (2.1 GHz) at -13 dBm for False Positive ESN predictions. A grayed node represents knowledge of the observed value of that (random) variable with the value at each node specified by the corresponding range selected.}
\label{fig:bnm_2p1GHz_m5dBm_lstm_fp}
\end{figure}

To perform the causal inference analysis, we construct a Bayesian Network Model (BNM) using historical ground truth (short or long-term depending on the expected timescales of change in joint input features-labels distribution) which describes the causal relationships among the system parameters of interest. From 3GPP technical specifications (TS) 38.214 ~\cite{std3gpp38214}, we know that the Channel Quality Indicator (CQI) and the uplink (UL) and the downlink (DL) Modulation and Coding Scheme (MCS) indices are the parameters that are directly or indirectly impacted by the presence of a jamming signal. 
The inter-dependence of these parameters can be represented in the BNM shown in Fig.~\ref{fig:bnm_2p1GHz_m13dBm}, which has been implemented in the Netica software tool and reflects the topology of Fig.~\ref{fig:causal}, with each node implementing a Conditional Probability Table (CPT). 
The statistical distributions of the data used to construct the CPT at each node in Fig.~\ref{fig:bnm_2p1GHz_m13dBm} is based on the data collected in our experiments.
% \textcolor{blue}{(Describe what the notation `RangeAtoB' means in the BNM).}
The notation `RangeAtoB' in the BNM diagrams of Fig.~\ref{fig:bnm_2p1GHz_m13dBm}, Fig.~\ref{fig:bnm_2p1GHz_m5dBm_lstm_fn} and Fig.~\ref{fig:bnm_2p1GHz_m5dBm_lstm_fp} represents all observed values $\{h\}$ of that KPI (node in the BNM) that satisfy $A \leq h < B$.
Note here that the presence of jamming has a direct effect on the PUSCH SNR as well as the CQI, while the latter two parameters indirectly impact the DL MCS and UL MCS distributions.

% Bayesian Network Model (BNM) diagrams

Primarily focusing on the ESN detection model under the 2.1 GHz (LTE DL) -13 dBm jamming scenario, we first examine their corresponding incorrect predictions, i.e., FPs and FNs. A similar analysis can be performed for the LSTM or the instantaneous detection model. For these incorrect predictions, we obtain the distributions of the PUSCH SNR, CQI, DL MCS and UL MCS. 
The mean values of distributions of each aforementioned parameter is selected in the BNM model of Fig.~\ref{fig:bnm_2p1GHz_m13dBm} to obtain the aposteriori probability of jamming being present, denoted as $\Pr(H_1|\mathbf{\Theta})$, where $\mathbf{\Theta}$ denotes the data observed in the test stage, i.e., $\mathbf{\Theta} := \{\mathbf{Y}_i \}_{i=1}^{N_{\text{test}}}$ for $\mathbf{Y}_{i} \in \mathbb{R}^{N_f \times T}$, i.e., the number of multi-feature sequences observed in the test stage is $N_{\text{test}}$.
% \textcolor{blue}{(Give a clear definition of $\mathbf{\Theta}$).}
% \begin{figure}[h]
% \centering
% \includegraphics[width=\textwidth]{FIGURES_FINAL/BNM/BNM_2p1GHz_m5dBm_ESN_FP+FN.PNG}
% \caption{Bayesian Network Model for interference in the LTE DL band (2.1 GHz) at -5 dBm for incorrect ESN predictions.}
% \label{fig:bnm_2p1GHz_m5dBm_esn_fpfn}
% \end{figure}
% For the instantaneous random forest model, the aposteriori probability of interference is calculated to be $47.6\%$. 
For e.g., the BNM can correct $62.5\%$ of FN erroneous predictions of the ESN with a confidence $\Pr(H_1|\mathbf{\Theta})=68.2\%$ as seen in Fig.~\ref{fig:bnm_2p1GHz_m5dBm_lstm_fn}.
Similarly, the BNM can correct $80\%$ of the FP erroneous predictions of the ESN with a confidence $\Pr(H_1|\mathbf{\Theta})=9.07\%$ or $\Pr(H_0|\mathbf{\Theta})=90.93\%$ as seen in Fig.~\ref{fig:bnm_2p1GHz_m5dBm_lstm_fp}.
Overall, we find that this BNM-based posterior analysis can mitigate $72.2\%$ of the incorrect (FN + FP) predictions of the ESN.
% Furthermore, through this analysis, we find that the BNM can correct $86.4\%$ of incorrect predictions with high confidence, whereas it can correct $68.2\%$ of the incorrect predictions with lower confidence.
% For the LSTM as well as the ESN models, this probability is found to be $47.1\%$.
% This result is in accordance with the expectation that for incorrect predictions made by the three models, the aposteriori probability of jamming would be close to $50\%$.  
To further illustrate this performance improvement, the jamming detection performance metrics before and after the improvement are summarized in Table~\ref{tab:performance_improvement_bayesian}. We can observe an increase of $5.66\%$ in the accuracy along with a $80.16\%$ reduction in the False Alarm Rate (FAR) of the decision results of the near real-time ESN model, thereby making the detection decisions more reliable for further mitigation strategies.
The posterior probabilities classified according to the type of the erroneous decision, along with the average values of the KPIs (features) for both erroneous decision types are summarized in Table~\ref{table:aposteriori_prob}.

% Note that the BNM also supports correct predictions by the ESN as evidenced by the values of $\Pr(\text{jamming}|\mathbf{\Theta})$ for the TP and TN cases.
% For e.g., the BNM provides decision support with high confidence for $74.4\%$ of TP cases, whereas the BNM provides decision support for all TN scenarios but with a lower confidence.
This is an important result indicating that a Bayesian network constructed using domain knowledge and statistical relationships amongst different variables can be used to mitigate the erroneous predictions of learning-based detection models that need to be lightweight or have reduced capacity owing to computational complexity and power consumption constraints.
Note that the Bayesian network model in this study is not intended to replace the ML-based detection model. It serves as an evaluation and validation for ML models, especially for use in scenarios without an easy-to-detect pattern that requires a large amount of data to achieve higher accuracy. In addition, it can also be used to correct a significant portion of the incorrect predictions that are labeled with low confidence in the detection results of the ML-based models.
Note that although the BNM causal inference approach is described here for determining a binary detection decision as a proof-of-concept, it can be extended to a multi-class scenario determination as well with appropriate changes to the BNM of Fig.~\ref{fig:bnm_2p1GHz_m13dBm}.
This possibility has been taken into account in the flowchart in Fig.~\ref{fig:flowchart} outlining the overall jamming detection and classification module.

% Note the data values in the columns for the PUSCH SNR, CQI, DL MCS and UL MCS in Table~\ref{table:aposteriori_prob} refer to their respective mean values for a specific prediction output of the detector.

\begin{table*}[h]
\centering
\caption{Posterior Jamming Probablities $\Pr(H_1|\mathbf{\Theta})$ from the BNM based on Incorrect Detection Results of the ESN}
\begin{tabular}{ccccccc}
\hline
\textbf{Model Type and Scenario}                 & \textbf{Prediction Type} & \textbf{PUSCH SNR (dB)} & \textbf{CQI} & \textbf{DL MCS} & \textbf{UL MCS} & $\Pr(H_1|\mathbf{\Theta})$ \\ \hline
% \multirow{4}{*}{LSTM} & TP only                  & 13.77             & 15.00           & 25.88          & 18.98          & 50.20\%                  \\
%                                     & TN only                  & 14.23             & 14.22       & 24.71          & 18.39           & 45.70\%                  \\
%                                     % & TP+TN                    & 14.084             & 14.619       & 25.36           & 18.833          & 47.10\%                  \\
%                                     & FP only                  & 14.37                & 15.00          & 26.10             & 17.93             & 45.70\%                      \\
%                                     & FN only                  & 13.88              & 15.00           & 25.90           & 19.36           & 50.20\%   
%                                      \\ \hline
%                                     % & FP+FN                    & 14.02              & 15           & 26.09           & 19.17           & 47.10\%                  \\ \hline

\multirow{2}{*}{Bidirectional RC (ESN) (2.1GHz at -13dBm)}  & FP only                  & 15.31              & 11.54        & 21.34           & 18.66           & 9.07\%                  \\
                                    & FN only                  & 12.66              & 15.00        & 25.88           & 19.07           & 68.20\%                  \\
                                    % & FP+FN                    & 13.80              & 13.25        & 21.22           & 19.17           & 36.90\%                  \\ 
                                    \hline
\end{tabular}
\label{table:aposteriori_prob}
\end{table*}

\begin{table*}[!h]
\centering
\caption{Performance Improvement of the Bidirectional RC (ESN)-based Detector through Bayesian Causal Inference}
\label{tab:my-table}
\begin{tabular}{@{}cccccc@{}}
\toprule
\textbf{Case}        & \textbf{Accuracy (\%)} & \textbf{Precision (\%)} & \textbf{Recall (\%)} & \textbf{F-score (\%)} & \textbf{False Alarm Rate (\%)} \\ \midrule
Original ESN result  & 92.7419                & 92.1260                 & 93.6000              & 92.8571                 & 8.1301                         \\
After BNM correction & 97.9879                & 98.3806                 & 97.5904              & 97.9839                 & 1.6129                         \\ \bottomrule
\end{tabular}
\label{tab:performance_improvement_bayesian}
\end{table*}

% \textcolor{red}{Adherence to 3GPP standards:}
% Show that in the a posteriori BNM with SNR selected, the CQI and DL/UL MCS distributions are according to Tables in 38.214. 
Lastly, we corroborate the distributions of the CQI and the DL MCS values for a particular measured PUSCH SNR with the values suggested by 3GPP standards, specifically TS 38.214~\cite{std3gpp38214}. In particular, for a PUSCH SNR of $15$ dB, the spectral efficiency is $\eta=5.0278$ bits/sec/Hz. Mapping this to Table 5.1.3.1-1 from~\cite{std3gpp38214}, we can see that this spectral efficiency corresponds to a DL MCS index of $26$. Additionally, from Table 5.2.2.1-2 in~\cite{std3gpp38214}, the same spectral efficiency also corresponds to a CQI index in the range $14-15$. This inference holds true in the collected data across all jamming scenarios and can also be seen in the distributions of these parameters in the BNM of Fig.~\ref{fig:bnm_2p1GHz_m13dBm}, providing further confidence in the ability of the Bayesian network constructed from this data to mitigate incorrect model predictions.

\section{Conclusion and Future Work}
\label{sec:conclusion}

In this paper, we introduce a Bayesian Inference-assisted machine learning approach for jamming detection and classification in 5G New Radio (NR) networks, which leverages critical cross-layer signaling data collected on a real 5G NR NSA testbed and is calibrated via Bayesian Network Model (BNM)-based inference. 
The introduced approach leverages both instantaneous and sequential time-series data samples, achieving high accuracy and robust detection for known jamming types, and can address potentially low detection accuracy caused by limited training data by employing Bayesian causal analysis and inference. Our methodology connects BNM-based causal inference with the domain knowledge available in 3GPP technical specifications to detect and identify the root cause of any observed performance degradation and serves to validate as well as enhance the resilience of ML-based jamming detection models, where we successfully address 72.2\% of erroneous predictions by sequential models caused by insufficient training data. The introduced methodology is applicable for deployment in infrastructure Radio Access Network (RAN), network edge as well as UEs in 5G NR as well as B5G cellular networks. As part of future work, we plan to extend the introduced approach to unknown jamming waveforms using unsupervised and reinforcement learning-based methods.

\bibliographystyle{IEEEtran}
% \bibliography{IEEEabrv,references}
\bibliography{IEEEabrv,ref}

\end{document}